\addtocontents{}{}
\documentclass[11pt]{article}
\usepackage{amsmath,amssymb,color,graphics,epsfig,cite,footnote}

\usepackage{amsfonts}

\newcommand{\hoch}[1]{$\, ^{#1}$}

\newcommand{\be}{\begin{equation}}
\newcommand{\ee}{\end{equation}}
\newcommand{\bea}{\setlength\arraycolsep{2pt} \begin{eqnarray}}
\newcommand{\eea}{\end{eqnarray}}
\newcommand{\nn}{\nonumber}

\def\ft#1#2{{\textstyle{\frac{\scriptstyle #1}{\scriptstyle #2} } }}
\def\fft#1#2{{\frac{#1}{#2}}}

\def\0{{\sst{(0)}}}
\def\1{{\sst{(1)}}}
\def\2{{\sst{(2)}}}
\def\3{{\sst{(3)}}}
\def\4{{\sst{(4)}}}
\def\5{{\sst{(5)}}}
\def\6{{\sst{(6)}}}
\def\7{{\sst{(7)}}}
\def\8{{\sst{(8)}}}
\def\sst#1{{\scriptscriptstyle #1}}

\begin{document}

	
	\begin{center}
		{\Large {\bf Trapping Horizons of the Evolving Charged Wormhole and Black Bounce}}

		\vspace{20pt}
		
	{Jinbo Yang\hoch{1} and Hyat Huang\hoch{2}}

\vspace{10pt}

\hoch{1}{\it Institute for Theoretical Physics,\\
	 Kanazawa University, Kanazawa 920-1192, Japan}
	 \\
	 
\hoch{2}{\it College of Physics and Communication Electronics, \\
	Jiangxi Normal University, Nanchang 330022, China}

		\vspace{40pt}
		
		\underline{ABSTRACT}
	\end{center}
We obtain one family of dynamic solutions in the Einstein-Maxwell-scalar(EMS) theory.
Our solutions could describe the evolving charged black(white) hole or wormhole and its transition, including the case of black bounce/wormhole transition.  
We compare different wormhole throat definitions and suggest that the usage of trapping horizons is the most suitable choice for tracking the evolution of the dynamic black(white) hole and wormhole, and their conversion in a unified framework.
Then we research several evolving processes in the appropriate parameters region, including the charge, the initial condition for the scalar hair, and parameters in our EMS Lagrangian.
The results show that the appearance of a degenerate marginal trapped surface is the crucial event for the conversion or transition, particularly in these cases: i) when the evolving wormhole converts to a black hole, the surface emerges and splits into two trapping horizons; 
ii) if the metric would become a black hole but finally fails, two trapping horizons combine as the surface and then vanish; 
iii) if the black bounce/wormhole transition happens, one single trapping horizon changes its type.

\vfill {\footnotesize  j\_yang@hep.s.kanazawa-u.ac.jp\ \ \ hyat@mail.bnu.edu.cn}

	\thispagestyle{empty}
	
	\pagebreak
	\tableofcontents
\addtocontents{toc}{\protect\setcounter{tocdepth}{2}}
	


\section{Introduction}\label{se1}
Black holes and wormholes are two kinds of fascinating objects predicted by general relativity.
Recent observations provide strong evidence for the presence of black holes\cite{Abbott:2016blz, TheLIGOScientific:2016src, Akiyama:2019cqa}.
As a massive object characterized by its horizon, a black hole is a robust prediction of general relativity that plays crucial roles in several fields about astrophysics, cosmology, and the frontier of theoretical physics\cite{Penrose:1964wq, Rees:1984si, Almheiri:2012rt}.
On the other hand, though the presence of wormholes is still less evident, it is an interesting hypothesis prediction from general relativity. 
The concept of the wormhole can be traced back to the paper written by Flamm in 1916\cite{flamm}. Later in 1935, Einstein and Rosen proposed a ``tunnel" geometry that connects two asymptotic Minkowski worlds without an explicit singularity\cite{Einstein:1935tc}. 
Morris and Thorne revived the investigations for traversable wormholes in 1988\cite{Morris:1988cz, Morris:1988tu}.
They showed that maintaining a traversable wormhole needs exotic matter that violates the null energy condition (NEC) according to classical general relativity.
Further, they also studied how to build a time machine using a ``short-cut" type traversable wormhole, which connects separate regions of the same universe.
The metrics they used are called Morris-Thorne wormholes.

The simplest exact traversable wormhole solution was suggested by Ellis in 1973, earlier than the Morris-Thorne wormhole\cite{Ellis:1973yv}.
It is a solution for the Einstein-scalar theory in which the scalar is the phantom field. This scalar field has a sign-flipped kinetic term in the Lagrangian, such that it plays the role of exotic matter.
Although phantom fields may cause instability problems due to their not-bound-from-below energy, theories with phantom fields could appear as some kinds of effective theories\cite{Carroll:2003st, Nojiri:2003vn}.
Bronnikov has also obtained the same static wormhole solutions in the same year\footnote{Then the solution is called Ellis-Bronnikove wormhole, including symmetric and asymmetric situations.}\cite{Bronnikov:1973fh}.
Further, Ellis also found the evolving version of his solution\cite{Ellis:1979bh}.
And it is very natural for finding the charged or rotating  wormhole solutions\cite{Nozawa:2020wet,Chew:2018vjp,Chew:2019lsa}.
Some researchers even proof the unique theorem for wormhole solutions in the Einstein-Maxwell-scalar (EMS) theory\cite{Lazov:2017tjs, Lazov:2019bni}.
Meanwhile, it is also possible to avoid the requirement of exotic matter in the context of modified gravity\cite{Mai:2017riq, Canate:2019spb}.

In the recent decade, theoretical progress strengthened the connection between black holes and wormholes.
The firewall argument sharpens the contradiction between the principles of quantum mechanics and general relativity\cite{Almheiri:2012rt}. It inspires the ER=EPR conjecture, which expects that two particles in the Einstein-Podolsky-Rosen (EPR) state keep in connection through a microscopic Einstein-Rosen (ER) bridge\cite{Maldacena:2013xja}.
Later, it is suggested that quantum teleportation corresponds to the traversable wormhole in the context of ER=EPR\cite{Gao:2016bin}.
Following this approach, the research for traversable wormholes is revived again\cite{Maldacena:2020sxe, Cubrovic:2020iad}.
It is worth noting that recent progress tells us constructing wormhole solutions without exotic matter or modified gravity is possible. 
Though the traversable wormhole violates the NEC, one crucial observation is that fermions offer negative Casimir-like energy. Thus it could keep a traversable wormhole open\cite{Maldacena:2018gjk}.
Some solutions were obtained in the Einstein-Dirac theory or the Einstein-Maxwell-Dirac theory \cite{Blazquez-Salcedo:2019uqq, Blazquez-Salcedo:2020czn}.

Simpson and Visser have suggested one kind of metric which connects different situations. Adjusting the parameters, the metric could become a Schwarzschild black hole, a regular black hole with a bounce instead of a singularity, or a traversable wormhole\cite{Simpson:2018tsi}.
They call the case of a bounce that happens behind the black hole horizon black bounce.
Bronnikov has also suggested a similar idea called black universe\cite{Bronnikov:2006fu}.
The investigations done by Bronnikov are motivated by studying regular black holes.
The singularity behind the horizon is avoided.
In some quantum gravity theories, like loop quantum gravity, not only the singularity in the early universe but also the Schwarzschild singularity could be avoided by quantum geometric effect\cite{Ashtekar:2020ifw}.
The black bounce metric given by Simpson and Visser gives further interesting scenarios.
It may be hard to distinguish a black hole and a wormhole just from the outside.
To clarify this, much research for lensing, quasinormal modes, echoes, etc. in the background of the Simpson-Visser or other wormhole metrics were studied\cite{Tsukamoto:2020bjm, Cheng:2021hoc, Islam:2021ful, Churilova:2019cyt, Bronnikov:2019sbx, Junior:2020lse,Blazquez-Salcedo:2018ipc,Liu:2020qia,Bronnikov:2021liv}. 
Further, Simpson and Visser have introduced Vaidya mass to their black bounce/wormhole metrics\cite{Simpson:2019cer}.
One can trace the black bounce to the wormhole, or the wormhole to the black bounce, transition process in detail by this dynamic Simpson-Visser metric. 

S.A.Hayward has also suggested a similar idea about black hole/wormhole transitions\cite{Hayward:1998pp,Hayward:2009yw,Shinkai:2002gv}.
This work was motivated by the study of the trapping horizon for the black hole.
It is well known that the event horizon is the boundary of a black hole,
but the event horizon is too foreseeing \cite{Mcnutt:2021qch}.
Despite that the event horizon is the Killing horizon for a stationary black hole,
it is hard to say where the horizon is just using the data in a particular moment.
Especially for numerical relativity, researchers prefer the quasilocal concept, apparent horizon, as the boundary of a black hole \cite{Cai:2008mh, Fan:2016yqv, Huang:2019lsl}.
Otherwise, one must use the information of the final state of the spacetime to determine the location of the event horizon.
The apparent horizon is where the expansion of outgoing null geodesics vanishes. 
Hayward defined the trapping horizon by this same property and classified it into different types\cite{Hayward:1993wb, Hayward:1994bu, Hayward:1997jp, Hayward:2005gi, Hayward:1998pp}.
He also found that this concept is also suitable for describing a wormhole.
Though the wormhole throat is well defined in a stationary situation, how to define a wormhole throat in general is still an open problem \cite{Hochberg:1998ii, Hayward:1998pp, Maeda:2009tk, Tomikawa:2015swa, Bittencourt:2017yxq, Mcnutt:2021qch}.
The earliest suggestion is from the Hochberg-Visser definition\cite{Hochberg:1998ii}.
Later, Hayward suggested it is sufficient to say there is a wormhole throat if a trapped or antitrapped region bound by two temporal trapping horizons exists\cite{Hayward:1998pp}.
Other definitions like the Maeda-Harada-Carr and the Tomikawa-Izumi-Shiro wormhole throats were also suggested\cite{Maeda:2009tk, Tomikawa:2015swa}.
We will discuss them in Sec. III.

In previous work \cite{Huang:2019arj}, we generalized the static Ellis wormhole to a static metric, which serves as a charged wormhole or a regular charged black hole depending on the scalar hair and charge. 
The new metric is a solution in the theory which included an additional Maxwell field nonminimally coupled to the free scalar field. In this paper, we find the dynamic version of this metric following the same EMS theory, and realize the dynamic black hole/ wormhole transition processes which are similar to the suggestion proposed by Simpson and Visser \cite{Simpson:2019cer}.
However, different from their designed dynamic metric, the size of the wormhole/bounce and the Vaidya mass are related to each other in our dynamic solution. One could not change the Vaidya mass without changing the size and vice versa. We should note that the scalar field and the Maxwell field in our theory could be both phantomlike and lead to violating the null energy condition.

With the different values of the parameters in the theory, our dynamic solution describes two kinds of physical processes of gravitational collapse. One situation shows the evolution end to the charged Ellis wormhole or black bounce in Ref.\cite{Huang:2019arj}. Another situation is eternal evolution. In some situations of eternal evolution, the final fate of the spacetime is similar to the dynamic Ellis wormhole in the far future. The beginning of the evolution could be a null singularity or look like a dynamic Ellis wormhole in the far past. 
It is worth noting that our dynamic metric would probably run into a medium phase that could become a black hole but finally fail.
We introduce the concept of a quasi-black hole to describe these situations.

The paper is organized as follows.
In Sec. II we will review our Einstein-Maxwell-scalar theory and give the dynamic solutions. Then we calculate the Vaidya masses of the dynamic solutions. We also ensure the solutions could reduce to the static solutions in Ref.\cite{Huang:2019arj}. 
In Sec. III we will give more details of the concept of trapping horizons. 
Then we will use the dynamic Ellis wormhole as a simple enough example to compare four definitions for the dynamic wormhole throat.
We will also explain why we chose Hayward's trapping horizon for describing the evolution of our dynamic black holes or wormholes.
In Sec. IV we will display how trapping horizons evolve in different cases, including the dynamic black hole/wormhole transition. 
Finally, we give a summary and discuss other open problems in Sec. V.

\section{The theory and solution}\label{sec2}

We consider an EMS theory in this work. The Lagrangian of the theory takes the form
\be\label{lag}
{\cal L}=\sqrt{-g}(R+\ft 1 2 (\partial \phi)^2-\ft{1}{4Z}  F^{2}),
\ee
where
\bea
Z(\phi)&=&\gamma_1\cos\phi+\gamma_2\sin\phi
\eea
is the coupling function of the scalar field and the Maxwell field is $F=dA$. Note that the scalar field is a phantom scalar because of its sign-flipped kinetic term. The Maxwell field could also become phantomlike due to the form of the coupling function $Z$. We investigated the same Lagrangian with \eqref{lag} in Ref.\cite{Huang:2019arj}, where we obtained static traversable wormhole and regular black hole solutions. 

The equations of motion of $g_{\mu\nu}$, the Maxwell field $A_\mu$, and the phantom scalar $\phi$ are respectively given by
\bea
&&\Box\phi =-\fft 14 \fft{\partial Z^{-1}}{\partial \phi}F^2 \,,\cr
&&\nabla_\mu\big(Z^{-1}F^{\mu\nu}\big) = 0\,, \cr
&&E_{\mu\nu} \equiv R_{\mu\nu}-\ft 12 R g_{\mu\nu}-T_{\mu\nu}^{A}-T_{\mu\nu}^{\phi}=0\,,
\eea
where
\bea
T_{\mu\nu}^{\phi} &=&-\bigg(\ft 12\partial_\mu\phi \partial_\nu \phi-\ft 14 g_{\mu\nu} (\partial \phi)^2\bigg),\nn\\
T_{\mu\nu}^{A} &=&\ft 12 Z^{-1}\Big(F_{\mu\nu}^2-\ft 14 g_{\mu\nu} F^2 \Big) \,.\\
\eea
For the spherically symmetric metric, we obtain a dynamic solution, which is given by 
\bea\label{sol2}
ds^{2} &=& -h dv^{2}+2 dr dv  +(r^{2}+a^2)d\Omega_{2}^{2},\nn\\
h&=&1-\ft{Q^2\gamma_2 r}{4a(r^2+a^2)}+\ft{Q^2\gamma_1}{4(r^2+a^2)},\nn\\
\phi&=&2\arcsin(\ft{a}{\sqrt{r^2+a^2}}),\quad A = \xi(r,q,Q,v) dv , \quad \xi'=\ft{QZ}{r^2+a^2}.
\eea
where $Q$ is a constant which corresponds to the electric charge.
In the viewpoint of $r\to\pm\infty$ observers, the Vaidya mass of the solution is given by
\be\label{M}
M_+= \ft{\gamma_2 Q^2}{8a}+a\dot{a}, \qquad\qquad M_-=-\ft{\gamma_2 Q^2}{8a}+a\dot{a}.
\ee
It is not a trivial work to read the mass from the metric even in the spacetime of the stationary wormhole \cite{Huang:2020qmn,Bronnikov:2021liv}. We have chosen to expand by the areal radius $\sqrt{r^2+a^2}$ rather than $r$. 
We should note that the two masses satisfy $M_++M_-=0$ in the static limit. 
This is different from the designed geometry considered in Refs.\cite{Simpson:2018tsi, Simpson:2019cer}, the mass of this wormhole associates with the parameter $a$ which characterizes the size of the wormhole or bounce.

The time-dependent function $a=a(v)$ is governed by the second-order evolution equation
\be\label{evo}
Q^2\gamma_2 \dot{a}+8a^3 \ddot{a}=0.
\ee
This second-order evolution equation \eqref{evo} can be integrated once and leads to the first-order equation
\be\label{evo2}
\dot{a}=\ft{Q^2\gamma_2}{16 a^2}+C,
\ee
as we will see later, where $C$ is a nontrivial constant.  For the theory with $\gamma_2=0$, we have $a(v)=Cv+a_0$.
In the case of $C=0$ it just gives the static solution. If $C$ is a nonzero constant, we could adjust $v\rightarrow v-a_0/C$ such that there is $a=Cv$.

The dynamic solution is consistent with the static solution we found in Ref.~\cite{Huang:2019arj} when $\dot{a}=0$ and $a$ is replaced by $q$.
\bea\label{ax}
&&ds^2=-h dt^2+h^{-1}dr^2+(r^2+q^2)d\Omega_{2}^2\quad \phi=\phi(r,q), \quad A=\xi(r,q,Q)dt,\nn\\
&&h=1-\ft{\gamma_2 Q^2 r}{4q(r^2+q^2)}+\ft{\gamma_1 Q^2}{4 (r^2+q^2)},\qquad \phi=2\arccos(\ft{r}{\sqrt{r^2+q^2}}),\qquad \xi'=\ft{QZ}{r^2+q^2}.
\eea
These static solutions, which were first constructed in Ref.\cite{Huang:2019arj}, could describe a charged Ellis wormhole or a regular black hole including black bounce. Thus our dynamic solutions \eqref{sol2} are the dynamic generalizations of the static solutions \eqref{ax} and hence they could describe a dynamic wormhole or black hole. However, due to the evolution equation \eqref{evo} or \eqref{evo2}, we will find that dynamic solutions \eqref{sol2} do not always take the static solutions \eqref{ax} as the final states when the time $v$ goes to infinity. This novel feature implies that the evolution of a massive object described by our solution strongly depends on the initial value. We will show this in Sec. IV.

Before moving to the next section, we prefer to make some conventions. Noting that our solutions are symmetric under flipping the sign of $v$, $r$, and $\gamma_2$, we see that situations of $\gamma_2\leq0$ are just the $v$ and $r$ reversed version of $\gamma_2\geq0$.
Thus we always set $\gamma_2\geq0$, only keep interested with the case of $g_{vr}=1$, and stand with the observers at the infinity far region of the $r>0$ side throughout this paper.

\section{Trapping horizon, apparent horizon and wormhole throat}

In this section, we will discuss some characteristic surfaces for a black hole (white hole) and wormhole before studying the evolution of $a(v)$. 
The boundary for a static black hole is the event horizon.
Events behind or just lying on the event horizon cannot be observed by any distant observer, so the event horizon is a null hypersurface.
The area of the event horizon for a settled-down black hole does not change with time.
Meanwhile, one of the most important surfaces for a stationary wormhole is the wormhole throat, which has the minimum area and satisfies the flare out condition. If a wormhole is traversable it means that its throat is a timelike hypersurface in spacetime. Therefore, this hypersurface can be foliated by a series of extreme surfaces with the same area. The timelike feature of the wormhole throat distinguishes from the event horizon of a stationary black hole, though both of their areas do not change with time.

However, both the wormhole throat and the event horizon are obscure in dynamic solutions.  
Since the event horizon is defined globally, one has to solve the dynamic black hole metric first, then find the event horizon by the whole information of the metric.
On the other hand, there is no reason to expect that any extreme surface exists for a dynamic wormhole.
Ashtekar, Hayward, etc. suggested using a quasilocal definition for the black hole horizon\cite{Ashtekar:2004cn, Hayward:1993wb}. 
Most of them use the same main property: a vanishing expansion for null geodesics.
Furthermore, Hayward has found that one of the quasilocal definitions for the black hole horizon is also suitable for defining the dynamic wormhole throat, such that the quasilocal horizon of a black hole and the dynamic wormhole throat share the same local features.

\subsection{Marginal surface and trapping horizon}
It would be useful to clarify the concepts of untrapped, (anti) trapped, and marginal surface first. 
Suppose we have an orientable codimension 2 closed surface, $\mathcal{S}$. 
Take two sets of the future-pointed null vector fields $m^{\mu}$ and $n^{\mu}$, which are orthogonal to this surface with the normalization $m^{\mu}n_{\mu}=-1$.
Then, the reduced inverse metric of the surface $\mathcal{S}$ is 
\be
h^{\mu\nu} = g^{\mu\nu} + m^{\mu}n^{\nu} + n^{\mu}m^{\nu}.
\ee
The expansions of $m^{\mu}$ and $n^{\mu}$ are 
\be
\theta_{(m)}=h^{\mu\nu}\nabla_{\mu}m_{\nu} \,,\quad \theta_{(n)}=h^{\mu\nu}\nabla_{\mu}n_{\nu} .
\ee
A surface $\mathcal{S}$ is called untrapped, if $\theta_{(m)}$ and $\theta_{(n)}$ have opposite signs.
If both expansions are negative or positive, then $\mathcal{S}$ is a trapped or antitrapped surface.
The surface $\mathcal{S}$ is marginal, if $\theta_{(m)}=0$ or $\theta_{(n)}=0$.
\textit{The definition of the trapping horizon is the hypersurface foliated by marginal surfaces.}

Without loss of generality, we assume it is the $\theta_{(n)}$ that vanishes at the marginal surface.
Based on the behaviors of another expansion $\theta_{(m)}$ and the Lie derivative $\mathcal{L}_m\theta_{(n)}$, we could further classify different types of marginal surfaces, which are listed in Table I 
and draw an intuited picture in Fig. 1. 
We can use the same classification for trapping horizons since a trapping horizon could be treated as a world tube of the marginal surface\cite{Sherif:2019vvo, Raviteja:2020fzt}.
Reference \cite{Helou:2015zma} also makes a clear summarization for the four nondegenerate situations.
\begin{table}
	\centering
	\caption{Classification of marginal surfaces }
	\begin{tabular}{| c | c | c | c | c | }
		\hline  
		{\bf Type} & {\bf $\theta_{(m)}$}  &{\bf $\mathcal{L}_{m} \theta_{(n)}$} &{\bf Examples}  \\
		\hline
		{\bf Future Outer}(FO)& $<0$  & $<0$ &Black hole horizon, wormhole throat   \\ \hline
		{\bf Future Inner}(FI)& $<0$  & $>0$ &Cauchy horizon, anti-Hubble horizon, \\ \hline 
		{\bf Past Outer} (PO) & $>0$  & $<0$ &White hole horizon, wormhole throat    \\ \hline 
		{\bf Past Inner} (PI) & $>0$  & $>0$ &Past Cauchy horizon, Hubble horizon, \\ \hline 
		{\bf Future Degenerate} (FD)& $<0$  & $=0$ & Extreme black hole horizon  \\ \hline 
		{\bf Past Degenerate }(PD)  & $>0$  & $=0$ & Extreme white hole  horizon \\ \hline 
	\end{tabular}
	\label{defitinition}
\end{table}

\begin{figure}[h]
\centering
\includegraphics[width=10cm]{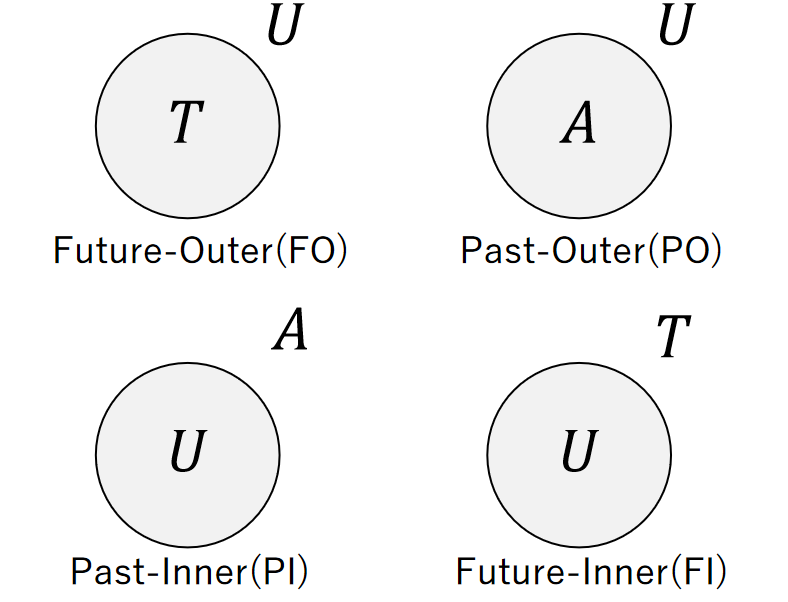}
\caption{\small Intuited picture for the classification of nondegenerate marginal surfaces.
T labels the trapped region;
A labels the antitrapped region;
U labels the untrapped region.
If an untrapped region surrounds the (anti) trapped region, the boundary is called outer.
The boundary for the opposite situation is called inner;
Future or past depends on trapped or antitrapped regions.}
\end{figure}

Since there is $\mathcal{L}_{m} \theta_{(n)}\neq 0$ for nondegenerate types, the covector field $\nabla_{\mu}\theta_{(n)}$ is nonvanishing.
This is the normal covector field for the trapping horizon $\theta_{(n)}=0$.
We could use the sign of $\mathcal{L}_{n} \theta_{(n)}$ to justify whether the normal vector field $\nabla^{\mu}\theta_{(n)}$ is timelike or spacelike, 
which shows the trapping horizon is a spatial or a temporal hypersurface, because of 
  \be
   \nabla^{\mu}\theta_{(n)} =- \mathcal{L}_{n} \theta_{(n)} m^{\mu} - \mathcal{L}_{m} \theta_{(n)} n^{\mu} \,.
  \ee
Meanwhile, according to the Raychaudhuri equation \cite{Raychaudhuri:1953yv}
  \be
   \mathcal{L}_{n} \theta_{(n)}=-\frac{1}{2}\theta_{(n)}^2 
                -\sigma_{\mu\nu}\sigma^{\mu\nu} +\omega_{\mu\nu}\omega^{\mu\nu} 
                -R_{\mu\nu}n^{\mu}n^{\nu}-\kappa\theta_{(n)},
  \ee
the sign of $\mathcal{L}_{n} \theta_{(n)}$ at the trapping horizon also shows whether the NEC is violated.
This is because the null congruence described by $n^{\mu}$ is orthogonal to the marginal surface such that $\omega_{\mu\nu}=0$ and $\theta_{(n)}=0$ at the trapping horizon. 
Also, $\sigma_{\mu\nu}\sigma^{\mu\nu}$ is positive so $\mathcal{L}_{n} \theta_{(n)}>0$ only if $R_{\mu\nu}n^{\mu}n^{\nu}<0$.
Especially in spherically symmetric spacetime, since there is $\sigma_{\mu\nu}=\omega_{\mu\nu}=0$ for radial null congruence, the sign of $R_{\mu\nu}n^{\mu}n^{\nu}$ directly determines the sign of $\mathcal{L}_{n} \theta_{(n)}$.
We can claim that the NEC is violated at the trapping horizon if the horizon is temporal outer or spatial inner.
Simultaneously, $\mathcal{L}_{n} \theta_{(n)}>0$ at the marginal surface also means that the area of the null geodesics congruence takes minimum. 
Namely, the flare out condition is satisfied.

\begin{table}
	\centering
	\caption{Spatial, Temporal, and Null Trapping Horizons }
	\begin{tabular}{| c | c | c | c | c |}
		\hline  
		{\bf Type} & {\bf$\mathcal{L}_{n} \theta_{(n)}$}  &{\bf $\mathcal{L}_{m} \theta_{(n)}$}& $\nabla^{\mu}\theta_{(n)}$ &{\bf Energy flux}  \\
		\hline
		{\bf Spatial Outer} & $<0$  & $<0$  &time-like  &Positive flux into FO or from PO  \\ \hline
		{\bf Temporal Outer}& $>0$  & $<0$  &space-like &Negative flux into FO or from PO \\ \hline
		{\bf Null Outer}& $=0$  & $<0$  &null &No flux  \\ \hline
		{\bf Spatial Inner} & $>0$  & $>0$  &time-like  &Negative flux into PI or from FI\\ \hline  
        {\bf Temporal Inner}& $<0$  & $>0$  &space-like &Positive flux into PI or from FI \\ \hline  
        {\bf Null Inner} & $=0$  & $>0$  &null  &No flux\\ \hline
	\end{tabular}
	\label{tab:epochs2}
\end{table}

It is worth noting that though one example of a degenerate trapping horizon is the extreme black hole or white hole horizon, we would see the emergence of a degenerate marginal surface signal trapping horizon pair creation(annihilation) or the transition between different types of trapping horizons in the latter discussion.

\subsection{Apparent horizon}
It is preferred to use the term apparent horizon to describe the evolution of a dynamic black hole. 
An apparent horizon is usually defined as a hypersurface with vanishing expansion for outgoing radial null geodesic congruences $n^{\mu}$ \cite{Cai:2006rs, Cai:2008mh, Fan:2016yqv, Huang:2019lsl},
\be
\theta_{(n)}=0.
\ee
This definition makes an apparent horizon exactly a trapping horizon.
The apparent horizon coincides with the event horizon for a stationary black hole. 
This is exactly the intuitional picture that an evaluating black hole finally settles down at a stationary phase.
As an example, we consider the Vaidya black hole. Its line element is
\be
ds^2=-(1-\ft{2m(v)}{r})dv^2+2dvdr^2+r^2d\Omega^2_2.
\ee
The apparent horizon of the Vaidya black hole is located at $r=2m(v)$. If there is a stationary point of the function $m(v)=m_f$, this metric will approximate to the Schwarzschild black hole with mass $m_f$ when the advanced time $v$ is large enough. The apparent horizon tends to and finally merges with the event horizon. 
For a white hole, one should consider the place where the ingoing null vector field has vanishing expansion as its boundary. The advanced time $v$ should be replaced by the retarded time $u$. The metric would have a minus $g_{ur}$ term. Settling down at the event horizon changes as beginning from the past event horizon.
More details and examples of the apparent horizon can be found in Refs.\cite{ Cai:2006rs, Cai:2008mh, Fan:2016yqv, Huang:2019lsl}.  
In this article, we use the term apparent horizon in the sense that the final state of hypersurface $\theta_{n}=0$ merges with the event horizon of a black hole, or the initial state of the hypersurface is from the past event horizon in the following section.

\subsection{Wormhole throats}

On the other hand, there is no agreed definition for the dynamic wormhole throat\cite{Hochberg:1998ii, Hayward:1998pp, Maeda:2009tk, Tomikawa:2015swa}. 
Reference \cite{Bittencourt:2017yxq} has compared several definitions in the circumstance of a wormhole in a Friedmann-Robertson-Walker universelike background.
We will give a short introduction here.
~\\

\noindent{\textit{Hochberg-Visser and Hayward}:}
Hochberg and Visser gave several definitions for the wormhole throat with different strengths \cite{Hochberg:1998ii}. 
The simplest one is the two hypersurfaces $\Sigma_{\pm}$, with vanishing expansion $\theta_{\pm}$ for affine parametric null vector field $l_{\pm}^{\mu}$, which satisfies the flare out condition $\frac{d\theta_{\pm}}{d u_{\pm}}\ge 0$ where the $u_{\pm}$ are affine parameters. 
The Hochberg-Visser(HV) definition may treat the bounce of the universe as a wormhole throat so in some circumstances the wormhole throat defined in this way does not match our intuition.    

Hayward defines the wormhole throats or mouths as temporal outer trapping horizons with mutual communication.
Later, in Ref.\cite{Hayward:1998pp}, Hayward gave a more flexible explanation about his definition.
Despite what a wormhole throat means, a wormhole region is bounded by two wormhole horizons which are locally like a black hole (FO) or white hole (PO) horizon but with negative energy.
Requiring mutual communication ensures the wormhole is traversable.
The wormhole horizon that satisfied Hayward's definition is also a wormhole throat in the sense of the HV definition.
~\\

\noindent{\textit{Maeda-Harada-Carr}:}
H. Maeda, T. Harada, and B. J. Carr (MHC) define the wormhole throat as having the minimum area on spacelike hypersurfaces\cite{Maeda:2009tk}.
It demands that we take a spacelike hypersurface as the whole space for a particular moment, 
while one could view the HV or Hayward definitions as having chosen null hypersurfaces. 
The MHC definition matches our intuition, but it depends on how we decompose the spacetime into a set of time slices when we discuss a dynamic wormhole.
~\\

\noindent{\textit{Tomikawa-Izumi-Shiromizu}:}
Y. Tomikawa, K. Izumi, and T. Shiromizu (TIS) define the wormhole throat as somewhere with satisfied $\theta_{k}=\theta_{l}$ \cite{Tomikawa:2015swa}.
The TIS definition is considered to be the hybrid between the HV/Hayward and the MHC definition.
Despite that it gets rid of which set of time slices we chose, this definition for a wormhole throat still depends on a chosen-by-hand structure.
If we rescale the null vector field like this:
\be
k^{\mu}\rightarrow \tilde{k}^{\mu}=ak^{\mu} \,,\,
l^{\mu}\rightarrow \tilde{l}^{\mu}=a^{-1}l^{\mu}
\ee
in which $a>0$ and $a\neq 1$. Since the vector fields $\tilde{k}^{\mu}$ and $\tilde{l}^{\mu}$ are still future pointed and satisfy the normalization $\tilde{k}^{\mu}\tilde{l}_{\mu}=-1$, the rescaling does not effect the HV and the Hayward definitions but shifts the hypersurface $\theta_{k}-\theta_{l}=0$ to 
\be
\theta_{\tilde{k}} - \theta_{\tilde{l}}
= a \theta_{k} - a^{-1}\theta_{l} =0 .
\ee

\noindent{\textit{Comparison by Evolving Ellis drain hole}:}
As a concrete example, we consider the metric which takes the following form
\be\label{dsdrdv}
ds^2=-hdv^2+2drdv+(r^2+a(v)^2)d\Omega^2_2.
\ee
The tangent vector fields of outgoing (denoted by $k^\mu$) and ingoing (denoted by $l^\mu$) radial null geodesic congruences are given by 
\bea
k^\mu\ft{\partial}{\partial x^\mu}&=&\ft{\partial}{\partial v }+\ft{h}{2}\ft{\partial}{\partial r},\\
 l^{\mu}\ft{\partial}{\partial x^{\mu}} 
 &=&  -\ft{\partial}{\partial r},
\eea
respectively.
The area of a hypersurface with constant $v$ and $r$ is given  by $A=4\pi (r^2+a^2)$. Thus the expansion of $k^\mu$ is 
\be
\theta_{k}=\ft{k^\mu \nabla_\mu A}{A}=\ft{r h +2a\dot{a} }{ r^2+a^2}.\label{nullk}
\ee
Similarly, the expansion of $l^\mu$ is given by 
\be\label{thetal}
\theta_{l}=\ft{l^\mu \nabla_\mu A}{A}=-\ft{2r}{r^2+a^2}.
\ee
A similar trick is also applied in Ref.\cite{Cai:2006rs}

If we set $Q=0$ for the solution given in Sec. II
, we obtain the metric \eqref{dsdrdv} where $h=1$ and $a(v)=Cv$. Thus the metric is the evolving flowless drain hole solution first proposed by H.Ellis in Ref.\cite{Ellis:1979bh}, Eq. (4.27), though our metric \eqref{dsdrdv} with $h=1$ looks different with the original line element 
  \be
   ds^2=-dt^2+dx^2+(\alpha^2 t^2 + (1+\alpha^2)x^2)d\Omega^2_2,
  \ee
We fix the expression in our convention. 
These two line elements are the same up to the coordinates transformation and parameter redefinition,
 \be
 t= \ft{(1+\alpha^2)v-r}{\sqrt{1+2\alpha^2}}\,,\, 
 x= \ft{\alpha^2v+r}{\sqrt{1+2\alpha^2}}\,,\,
 \alpha^2=\ft{\sqrt{1+4C^2}-1}{2} \, .
 \ee
Additionally, we redefine the two radial null vector fields,
\be
\tilde{k}^{\mu}\ft{\partial}{\partial x^{\mu}}=\ft{1}{\sqrt{2}}(\ft{\partial}{\partial t} + \ft{\partial}{\partial x}) \,,\, 
\tilde{l}^{\mu}\ft{\partial}{\partial x^{\mu}}
=\ft{1}{\sqrt{2}}(\ft{\partial}{\partial t} - \ft{\partial}{\partial x}) \,.
\ee
The new null vector fields rescale the fields $\{k^{\mu},l^{\mu} \}$ as
\be
\tilde{k}^{\mu}=\sqrt{\ft{2}{1+2\alpha^2}} k^{\mu}\;,\; 
\tilde{l}^{\mu}=\sqrt{\ft{1+2\alpha^2}{2}} l^{\mu} \;.
\ee
Their expansions are 
 \be
 \theta_{\tilde{k}}=\ft{\sqrt{2}(\alpha^2 t + (1+\alpha^2)x)}{\alpha^2 t^2 + (1+\alpha^2)x^2}\;,\;
 \theta_{\tilde{l}}=\ft{\sqrt{2}(\alpha^2 t - (1+\alpha^2)x)}{\alpha^2 t^2 + (1+\alpha^2)x^2} \;.
 \ee
 
Alternatively, we could choose another set of time slices,
 \be
  \bar{t}= v-r.
 \ee
Then the metric \eqref{dsdrdv} becomes
 \be
 ds^2=-d\bar{t}^2+dr^2+(r^2+C^2(\bar{t}+r)^2)d\Omega^2_2.
 \ee
For the following null vector fields:
  \be
  \bar{k}^{\mu}\ft{\partial}{\partial x^{\mu}}=\ft{1}{\sqrt{2}}(\ft{\partial}{\partial \bar{t}} + \ft{\partial}{\partial \bar{r} }) \,,\, 
  \bar{l}^{\mu}\ft{\partial}{\partial x^{\mu}}
  =\ft{1}{\sqrt{2}}(\ft{\partial}{\partial \bar{t}} - \ft{\partial}{\partial \bar{r} }) \,,
  \ee
their expansions are
  \be
  \theta_{\bar{k}}=\ft{\sqrt{2}( 2C^2\bar{t}+ (1+2C^2)r )}{ r^2+C^2(\bar{t}+r)^2 }\;,\;
  \theta_{\bar{l}}=-\ft{\sqrt{2}r}{r^2+C^2(\bar{t}+r)^2} \;.
  \ee
We should be careful about the fact that even though $\bar{r}=r$, the vector field $\frac{\partial}{\partial \bar{r} }$ in the new basis is different than the vector field $\frac{\partial}{\partial r }$.
Instead, it should be $ \frac{\partial v}{\partial \bar{r} } \frac{\partial}{\partial v } + \frac{\partial r}{\partial \bar{r} } \frac{\partial}{\partial r }
= \frac{\partial}{\partial r }- \frac{\partial}{\partial v } $.

We list the comparison in Table. III and Fig.2. Using the coordinate transformations, one could confirm that the choice of coordinates or scaling of null vectors is irrelevant for the wormhole throats under the HV/Hayward definition, while the MHC definition depends on time slices and the TIS definition depends on the scaling.

 \begin{table}
 	\centering
 	\caption{Comparing the definitions. }
 	\begin{tabular}{| c | c | c | c | c |}
 		\hline  
 		{\bf Coordinates} &{\bf Null Vectors} & {\bf HV/Hayward}  &{\bf MHC}& {\bf TIS}  \\
 		\hline
 		$\{v,r\}$, $v$ as time slices & $k^{\mu}$ and $l^{\mu} $ & $r=-2C^2v$ and $r=0$  & — & $r=-\frac{2C^2v}{3}$   \\ \hline
 		$\{\bar{t},r\}$, $\bar{t}$ as time slices & $\bar{k}^{\mu}$ and $\bar{l}^{\mu} $ & $r=-\ft{2C^2\bar{t}}{1+2C^2}$ and $r=0$  & $r=-\ft{C^2\bar{t}}{1+C^2}$  & $r=-\ft{C^2\bar{t}}{1+C^2}$   \\ \hline
 		$\{t,x\}$, $t$ as time slices & $\tilde{k}^{\mu}$ and $\tilde{l}^{\mu} $ & $x=-\frac{\alpha^2t}{1+\alpha^2} $ and $x=\frac{\alpha^2t}{1+\alpha^2}$  & $x=0$  & $x=0$   \\ \hline
 	\end{tabular}
 	\label{tab:epochs3}
 \end{table}

\begin{figure}[h]
\centering
\includegraphics[width=5cm]{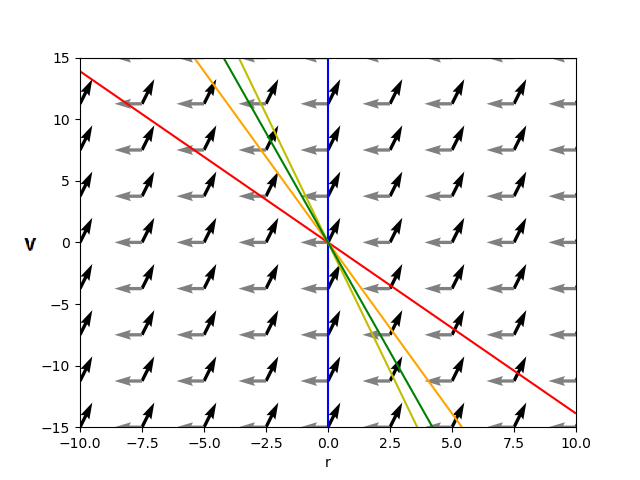}
\includegraphics[width=5cm]{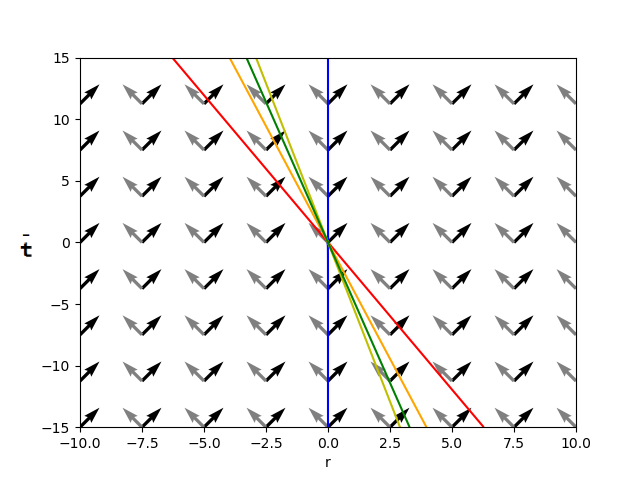}
\includegraphics[width=5cm]{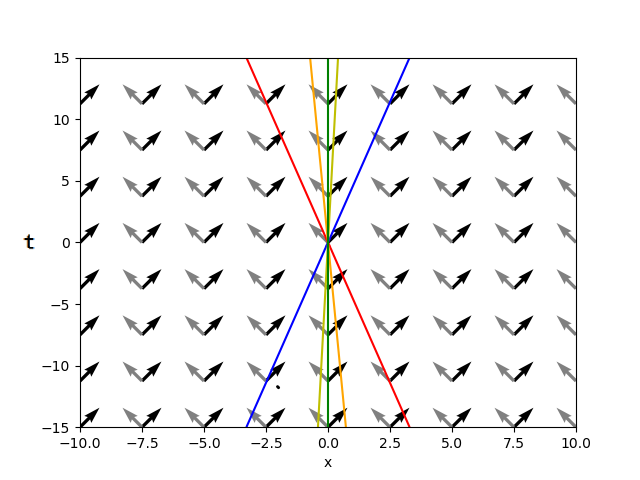}
\caption{\small Comparing different definitions for the wormhole throat.
       We set $C=0.6$ and draw the hypersurfaces in coordinates frames $\{v,r\}$, $\{ \bar{t},r\}$, and $\{t,x\}$. 
         The green line shows the MHC wormhole throat for $\tilde{t}$ which is the same as the TIS for $\{\tilde{k},\tilde{l} \}$;
         the orange line shows the MHC wormhole throat for $\bar{t}$ which is the same as the TIS for $\{\bar{k},\bar{l} \}$;
         the yellow line shows the TIS wormhole throat for $\{k,l \}$.}
         
\end{figure}

The geometry of the Ellis drain hole has a product structure $\mathcal{M}\times S^2$, in which the $\mathcal{M}$ is a two-dimensional Minkowski spacetime.
The areal radius $\rho$ can be viewed as a scalar field in  $\mathcal{M}$.
Thus, we could shift the MHC or TIS wormhole throat to any timelike hypersurface bounded by two wormhole horizons, namely the HV/Hayward wormhole throats. 

 Noting that the parameter $\gamma_1$ does not join the equation for $a(v)$ in Sec. II, the evolving function $a(v)$ could still be $a=Cv$ if $\gamma_2=0$,
 such that the solutions are still simple enough to allow us to draw spacetime diagrams in coordinates frames $\{v,r\}$, $\{ \bar{t},r\}$ and $\{t,x\}$.
 
 The line elements are
 
 \be
 ds^2=-d\bar{t}^2+dr^2 + (1-h)(d\bar{t}+dr)^2 + (r^2+C^2(\bar{t}+r)^2)d\Omega^2_2,
 \ee
 
 \be
   ds^2=-dt^2+dx^2 + (1-h)\ft{(dt+dx)^2 }{1+2\alpha^2} + (\alpha^2 t^2 + (1+\alpha^2)x^2)d\Omega^2_2,
 \ee
 
 and the null vector fields are
\be
\tilde{k}^{\mu}\ft{\partial}{\partial x^{\mu}}=\ft{2-h+2\alpha^2}{\sqrt{2}(1+2\alpha^2)}\ft{\partial}{\partial t} + \ft{h+2\alpha^2}{\sqrt{2}(1+2\alpha^2)}\ft{\partial}{\partial x} \,,\, 
\tilde{l}^{\mu}\ft{\partial}{\partial x^{\mu}}
=\ft{1}{\sqrt{2}}(\ft{\partial}{\partial t} - \ft{\partial}{\partial x}) \,,
\ee

\be
\bar{k}^{\mu}\ft{\partial}{\partial x^{\mu}}=\ft{2-h}{\sqrt{2}}\ft{\partial}{\partial \bar{t}} +\ft{h}{\sqrt{2}} \ft{\partial}{\partial \bar{r} } \,,\, 
\bar{l}^{\mu}\ft{\partial}{\partial x^{\mu}}
=\ft{1}{\sqrt{2}}(\ft{\partial}{\partial \bar{t}} - \ft{\partial}{\partial \bar{r} }) \,.
\ee

\begin{figure}[h]
\centering
\includegraphics[width=5cm]{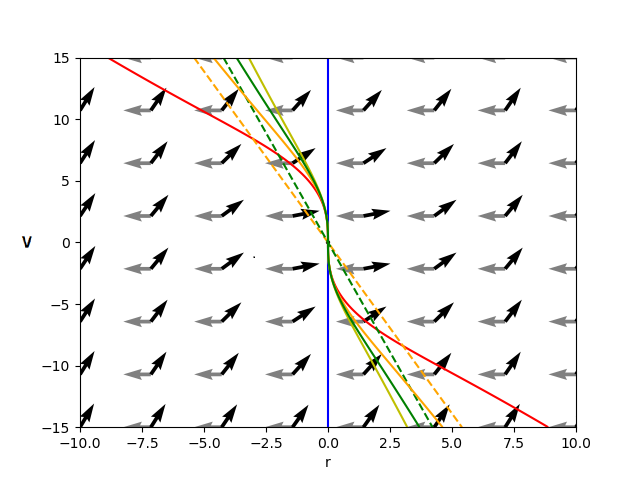}
\includegraphics[width=5cm]{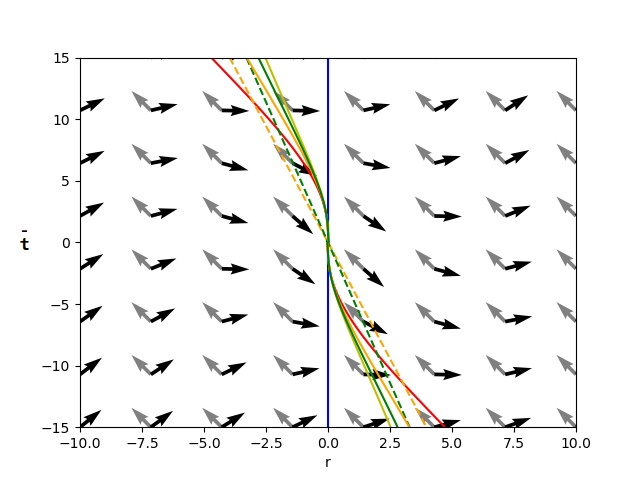}
\includegraphics[width=5cm]{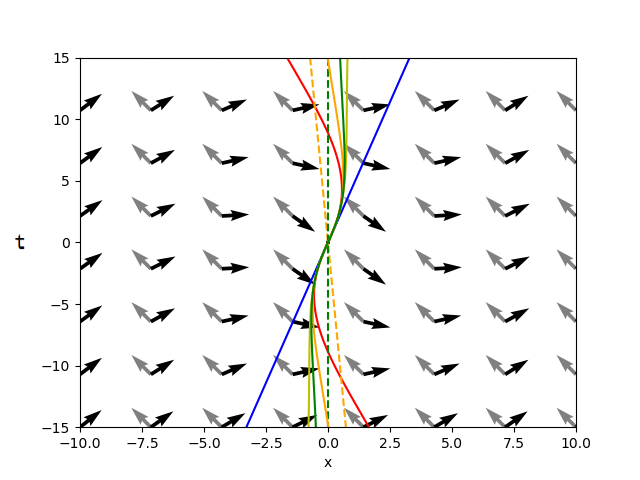}
\caption{\small
We set $Q=3.72$, $\gamma_1=10$, $C=0.6$.
The green line shows the TIS wormhole throat for $\{\tilde{k},\tilde{l} \}$;
the green dashed line shows the MHC throat for $\tilde{t}$;
the orange line shows the TIS throat for $\{\bar{k},\bar{l} \}$; 
the orange dashed line shows the MHC throat for $\bar{t}$;
the yellow line shows the TIS throat for $\{k,l \}$.}
\end{figure}

\begin{figure}[h]
\centering
\includegraphics[width=5cm]{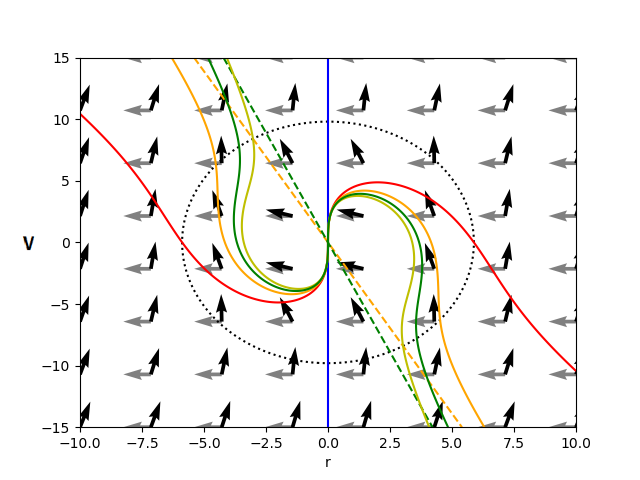}
\includegraphics[width=5cm]{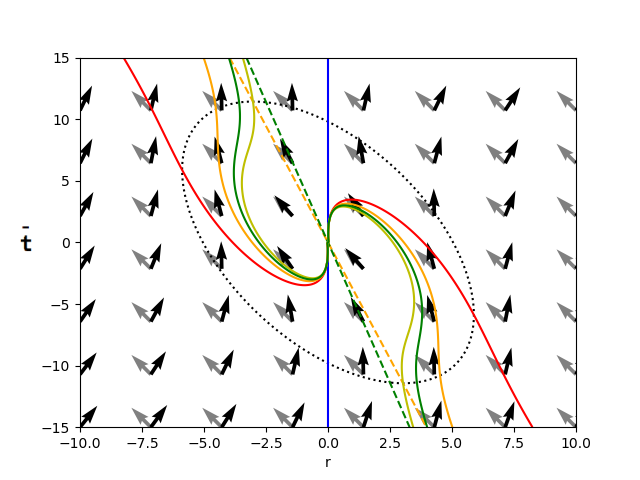}
\includegraphics[width=5cm]{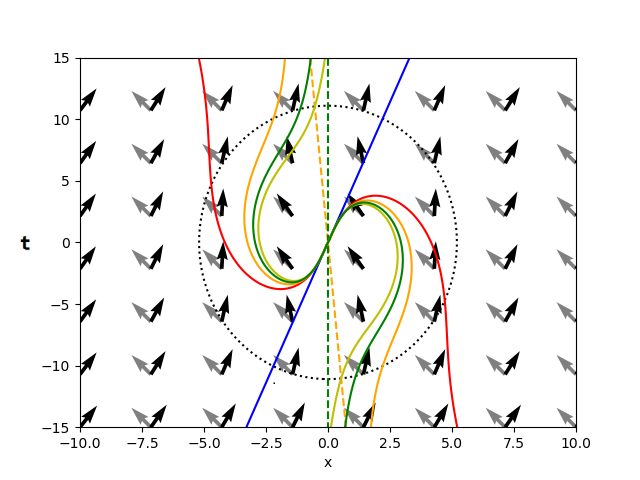}
\caption{\small
 We set $Q=3.72$, $\gamma_1=-10$, $C=0.6$.
 The green line shows the TIS wormhole throat for $\{\tilde{k},\tilde{l} \}$;
 the green dashed line shows the MHC throat for $\tilde{t}$;
 the orange line shows the TIS throat for $\{\bar{k},\bar{l} \}$; 
 the orange dashed line shows the MHC throat for $\bar{t}$;
 the yellow line shows the TIS throat for $\{k,l \}$;
 the black dotted line shows the hypersurface $h=0$.}
\end{figure}

 We calculate the corresponding expansion for the above null vector fields, then draw the wormhole throats under different definitions in Figs. 3 and 4.
 We found that the MHC wormhole throats for time slices $\bar{t}$ are different from the TIS wormhole throats for null vector fields $\{ \bar{k},\bar{l} \}$,
 while the MHC wormhole throats for time slices $\tilde{t}$ are different from the TIS wormhole throats for null vector fields $\{ \tilde{k},\tilde{l} \}$.

We would treat the HV/Hayward definition as the sufficient condition that the spacetime has wormhole throats.
More precisely, the picture we use is that the spacetime has a dynamic traversable wormhole if there are two temporal outer trapping horizons. 
But we do not directly treat any of these trapping horizons as wormhole throats.
We consider that the (anti) trapped region bounded by these two trapping horizons includes all wormhole throat candidates under the MHC definition or the TIS definition. 
We will focus on the evolution of temporal outer trapping horizons in the following section. 
It will be shown that, in some circumstances, the (anti) trapped region bounded by two temporal outer trapping horizons would tend to a minimum surface.
~\\

\noindent{\textit{ Bounce:}
The picture used by Hayward makes a more precise description for traversable wormholes since the HV definition also includes other situations like a bounce from a trapped region to an antitrapped region.
This inspires us to define bounce horizon in a similar way as Hayward for wormhole horizon. 
If there is an untrapped region bounded by two spatial inner trapping horizons, particularly past-bounded by future-inner trapping horizon (FITH) and future-bounded by past-inner trapping horizon (PITH), then we define these two trapping horizons as bounce horizons.
The requirement of spatial inner leads to the violation of the NEC, just as we summarized in Table.II.
For transition hypersurfaces, we could also make dual definitions with the MHC or TIS definitions for wormhole throats: a spatial hypersurface intersecting the minimum area of a series of timelike hypersurfaces or satisfying $\theta_k+\theta_l=0$.  
Also, the two spatial inner trapping horizons may tend to a spatial hypersurface foliated by minimum surfaces, which means $\theta_k=\theta_l=0$.
We will show all these circumstances that occur in our dynamic solutions in the next section.

We would also like to mention another interesting property here. Every situation of $\gamma_2=0$ includes an event singularity at the origin $v=r=0$. It represents when two future-outer trapping horizons (FOTH) combine and become two past-outer (POTH) if $\gamma_1\geq0$ (see Figs. 2 and 3), or when one PITH overlaps with one FITH if $\gamma_1<0$ (see Fig. 4).

\section{Black hole/wormhole transition}\label{4}

In the previous section, we explained why we chose trapping horizons to track the detailed evolution of a dynamic black hole or wormhole.
We would focus on the hypersurfaces given by the equations $\theta_{k}=0$ and $\theta_{l}=0$.
According to \eqref{thetal}, the equation $\theta_{l}=0$ always gives the root $r=0$. 
In most cases, this trapping horizon is outer, but it could be an inner trapping horizon in the case of $\gamma_1<0$.

The equation $\theta_{(k)}=0$ gives other trapping horizons.
We could replace $\dot{a}$ with $a(v)$ by using \eqref{evo2} to obtain a cubic equation from $\theta_{(k)}=0$, i.e,
\bea
&&a_1 r^3+b_1 r^2+c_1 r+d_1=0,\label{cubic}
\eea
in which 
\bea
&&a_1=8a,\nn\\
&&b_1=16C a^2-2\gamma_2Q^2,\nn\\
&&c_1=8a^3+2\gamma_1Q^2 a,\nn\\
&&d_1=16Ca^4+\gamma_2Q^2a^2.
\eea
We assume $\gamma_2 > 0$ since we have discussed the case of $\gamma_2=0$ in Sec. III. The cubic equation could have only one real root or have three real roots, which relate to one or three marginal surfaces concerning $\theta_k$ at a particular moment $v$. We define the discriminant for the real roots by 
\be
\Delta=(\ft{b_1 c_1}{6a_1^2}-\ft{b_1^3}{27 a_1^3}-\ft{d_1}{2a_1})^2+(\ft{c_1}{3a_1}-\ft{b_1^2}{9a_1^2})^3.
\ee
Equation \eqref{cubic} has three real roots if and only if $\Delta\leq 0$.

The first-order evolution equation \eqref{evo2} could be solved and rewritten as 
\bea
&&\text{case I :}   \dot{a}=\ft{Q^2\gamma_2}{16}\big( \ft{1}{a^2} - \ft{1}{a_0^2} \big),\label{evosluP}\\
&&\text{case II :}  \dot{a}=\ft{Q^2\gamma_2}{16}\big( \ft{1}{a^2} + \ft{1}{a_0^2} \big),\label{evosluS}\\
&&\text{case III :} \dot{a}=\ft{Q^2\gamma_2}{16 a^2},\label{evosluF}
\eea
where we use $a_0$ to replace $C$. 
It is straightforward to see that case I \eqref{evosluP} includes a stationary point $a=a_0$.
If the initial condition of $a$ is not $a_0$, case I shows that the evolution tends to $a=a_0$.
This means that when $v$ tends to plus infinity the spacetime would tend to a static solution obtained in Ref.\cite{Huang:2019arj}.
The final state would be a static black hole or a wormhole. 
However, case II \eqref{evosluS} and case III \eqref{evosluF} have no stationary point. These two cases depict an eternal evolution.

\subsection{Definitions and convention}

There are several situations during the dynamic processes in our solutions. To show the results distinctly, we will first clarify the definitions and our convention as follows.
~\\

\noindent{\textit{Null singularity:} 
The metric and curvatures are both singular while the scalar parameter $a$ goes to $0$. It is a null singularity that represents the beginning of evolutions in many cases. 
~\\

\noindent{\textit{Dynamic wormhole states:} 
We follow the suggestion given by Hayward\cite{Hayward:1998pp, Hayward:2009yw}.
During some era, the equation $\theta_k=0$ has only one root while the $r=0$ always corresponds to $\theta_l=0$,  such that the era includes two trapping horizons.
By using the Tables I and II, we could judge which type the horizons are.
If both of them are temporal outer, we say this era has a wormhole.
We call the moment $v$ a dynamic wormhole state if it cuts two temporal outer trapping horizons.
~\\

\noindent{\textit{Quasi-black hole states:} 
We could treat the FOTH as the boundary of a dynamic black hole.
It is consistent with the usual convention using the apparent horizon and the first motivation that Hayward introduced the trapping horizon\cite{ Hayward:1993wb, Hayward:1994bu, Hayward:1997jp, Cai:2006rs, Cai:2008mh, Fan:2016yqv, Huang:2019lsl}.
However, since the FOTH in our solution is temporal due to the ingoing phantom energy flux, it may also be one of the wormhole horizons as we've seen in Sec. III.
To distinguish between the dynamic wormhole states, we call the moment $v$ the quasi-black hole state if the moment $v$ does not cut another temporal outer trapping horizon near to the temporal FOTH.
The quasi-black hole state has four trapping horizons in most cases in our dynamic solutions\footnote{It would be more precise to also define quasi-white hole state if the relevant region is an antitrapped region rather than a trapped region, though this situation can be treated as a time-reversed quasi-black hole state.}.
On the other hand, our solutions also include the quasi-black hole that fails to finally form an event horizon.
~\\

\noindent{\textit{Static wormhole, RN-like black hole, black bounce:} 
These are the static solutions in Ref.\cite{Huang:2019arj}.
A static wormhole has a temporal hypersurface foliated by minimum surfaces.
The case of an RN-like black hole has further outer and inner Killing horizons,
while the case of black bounce has a spatial hypersurface foliated by minimum surfaces between its two Killing horizons.
We could treat the hypersurface foliated by minimum surfaces as two trapping horizons that coincide together.
In the following, we would see that the final fate of trapping horizons could be the outer or inner horizon of the black hole, or the hypersurface foliated by minimum surfaces.
~\\

In Table 
IV, we list the definitions of different regions for showing how trapping horizons divide the spacetime into untrapped, trapped, and antitrapped regions.
The untrapped regime is a spacetime regime that contains untrapped surfaces satisfying  $\theta_k>0$ and $\theta_l<0$. 
At the other side of $r$, there is another untrapped regime that contains surfaces, satisfying $\theta_k<0$ and $\theta_l>0$.
The trapped or antitrapped regime is the regime containing trapped or antitrapped surfaces.
\begin{table}[h]
 	\centering
 	\caption{Regions in spacetime }
 	\begin{tabular}{| c | c | c | c |}
 		\hline  
 		{\bf } &{\bf Label} & {$\theta_k$}  &{\bf $\theta_l$}  \\
 		\hline
 		Untrapped regions I & $U$  & $>0$  & $<0$ \\ \hline
 		Untrapped regions II& $U'$ & $<0$  & $>0$ \\ \hline
 		Trapped regions &$T$& $<0$ & $<0$   \\ \hline
	Antitrapped regions &$A$&  $>0$ & $>0$  \\ \hline
 	\end{tabular}
 	\label{tab:epochs4}
 \end{table}

\subsection{Evolution to a final stable state: Case I}

The first-order evolution equation of case I, i.e., \eqref{evosluP}, indicates that there is a stationary state at $a=a_0$. In fact, we could integrate \eqref{evosluP} directly and then obtain 
\be\label{case1}
\ln{\sqrt{|\ft{a+a_0}{a-a_0}|}}-\ft{a}{a_0} =\ft{Q^2\gamma_2}{16 a_0^3}(v-v_0),
\ee
where $v_0$ is a trivial integration constant which can be set to $0$. It is straightforward to see that the stationary point takes place at positive infinity of $v$ and hence our solution tends to a stationary metric as the final state.

The evolution equation \eqref{case1} has two branches depending on the initial condition. The $a-v$ plot in Fig.\ref{av1} shows that the lower branch starts from the null singularity $a=0$, increases monotonically, and finally tends to the $a=a_0$ state, while the upper branch evolves from a large $a$ state to the $a=a_0$ state smoothly. 
The $a(v)$ is monotonic in both two branches, which means that treating $a$ as a ``time" coordinate is possible.
For convenience, later we will use the $r-a$ plots to describe the detailed evolution of the spacetime and sketch the Penrose diagram for each branch. 
\begin{figure}[h]
\centering
\includegraphics[width=7cm]{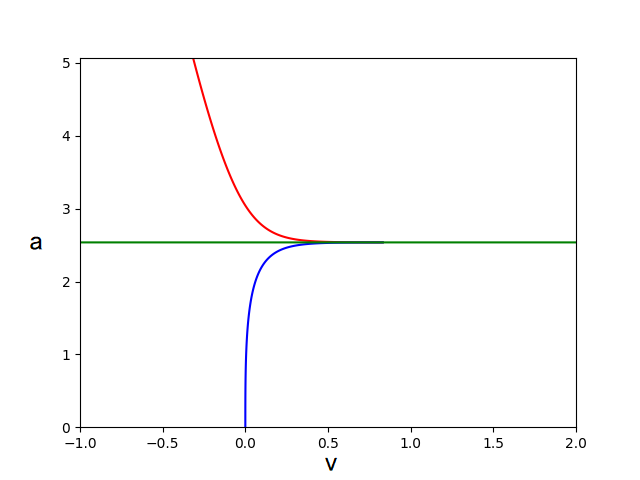}
\caption{\small  The two branches of $a(v)$ are shown in the same plot. The green line depicts the final value $a=a_0$, distinguishes the upper branch and lower branch. In the upper branch, the red line shows that $a(v)$ decreases to the green line ($a=a_0$) as time goes by. In the lower branch, the blue line shows that $a(v)$ increases to the green line ($a=a_0$) as time goes by. As we mentioned above, we could use $a$ as a new ``time" coordinate because of the monotonicity. We set $Q=3.72, \gamma_1=4.0, \gamma_2=5.3$ and $ a_0=2.54$.}\label{av1}
\end{figure}

The sign of the parameter $\gamma_1$ gives rise to different evolution. We would distinguish these situations in the following discussion.

\subsubsection{$\gamma_1>0$}

In the cases of $\gamma_1>0$, we will see the formation processes of a regular RN-like black hole and charged Ellis wormhole.

\noindent{\bf RN-like black hole:}

If the ratio $Q/a$ is more than the critical value  $8(\gamma_1+\sqrt{\gamma_1^2+\gamma_2^2})/\gamma_2^2$, the solution evolves to a final regular RN-like black hole. Figure \ref{RNra} depicts this situation.
The green line is the hypersurface $a=a_0$ corresponding to the final stationary state at positive infinity of $v$.
The black dotted line is the contour of $h=0$.
It surrounds the $h<0$ region with the $r$ axis, such that there is $h>0$ outside of the region.
The $h=0$ intersects with the $a=a_0$ at two locations corresponding to two Killing horizons of the static RN-like black hole, so we use the same labels $r_+$ and $r_-$ in which $r_+>r_->0$ to mark two roots of the equation $h=0$ when $a=a_0$.
The red line represents $\theta_{k}=0$ while the blue line represents $\theta_{l}=0$ which is always $r=0$.
The $\theta_{l}<0$ regime is at the right side of the $r=0$ and the $\theta_{l}>0$ is at the left side.
The $\theta_{k}=0$ also intersects with the $a=a_0$ at $r_+$ and $r_-$.
The rightmost part and the inside of the closed red line correspond to $\theta_{k}>0$, while another connected regime in the plot corresponds to $\theta_{k}<0$.
Thus, the whole plot is divided by the red and blue contours.
We could use Table IV 
to label every regime. 
Then we could judge the type and the nature of every trapping horizon by using Tables I and II.

Meanwhile, the streamlines are the integral curves for the null vector field $k^{\mu}$.
We hide the null vector field $l^{\mu}$ because it is just pointed to the left horizontally.
According to \eqref{nullk}, the field $k^{\mu}$ has the left component inside the $h<0$ region, becomes vertical at the hypersurface $h=0$, then obtains the right component outside.
This observation helps us to see where the streamlines turn to near the $a=a_0$.
It is shown that the streamlines which go through the far right red line would leave the $r_+$ and escape to infinity, i.e., $r\rightarrow\infty$. 
There exists one set of streamlines that hit $r_+$ exactly.
They form the event horizon. 
The other streamlines converge to the $r_-$.
Noting that the value of $a$ finally settles down at $a_0$, the convergence of those streamlines tells us their areas finally stop at the finite value $r_-^2+a_0^2$, apparently.
However, we should remind ourselves that the vector field $k^{\mu}$ would not lead to an affine parameter valued geodesic equation. Instead, the $k^{\mu}$ satisfies $k^{\nu}\nabla_{\nu}k^{\mu} = \ft{h'}{2}k^{\mu}$,
while there is $l^{\nu}\nabla_{\nu}l^{\mu}=0$ for the field $l^{\mu}$.
At the regime closed to the $r_-$, the spacetime is nearly the corresponding solution with the Killing vector field.
In Ref.\cite{Huang:2019arj}, we have drawn the Penrose diagrams for the stationary solution.
The diagrams show the hypersurface $r_-$ is a Killing horizon and plays the role of a Cauchy horizon in the situations of $r_-\neq r_+$.  
Thus, it would be more natural to expect that the spacetime could be extended and that the vector field $\ft{\partial}{\partial v}$ becomes a Killing vector field behind the $r_-$. 
We sketch the Penrose diagram within this consideration.
We still use the red lines to represent $\theta_{(k)}=0$ and the blue lines to represent $\theta_{(l)}=0$, the same with the $r-a$ plot.
Different from the $r-a$ plot, it is easy to identify the event horizon and we mark it as the yellow line.
The brown double line shows the null singularity while the dot-slash line shows the settle-down minimum surfaces, namely the wormhole throat in the stationary regime.

\begin{figure}[h]
\centering
\includegraphics[width=7cm]{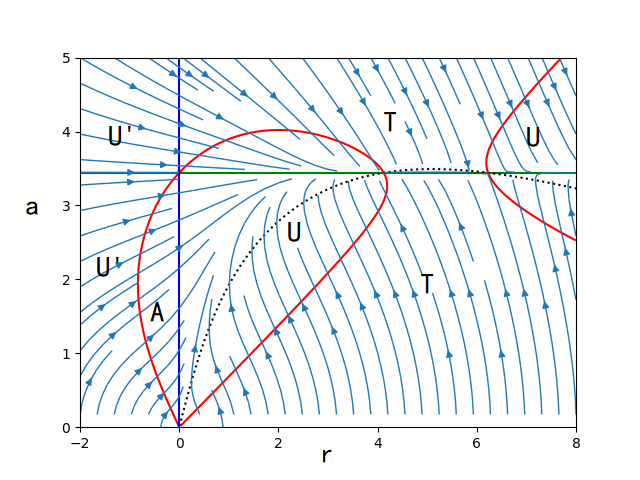}
\includegraphics[width=3.3cm]{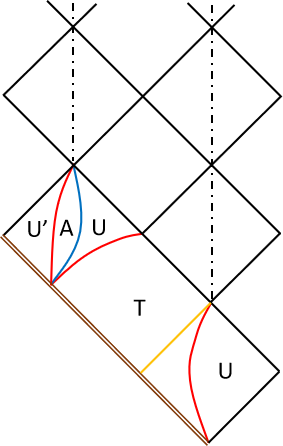}
\includegraphics[width=3.3cm]{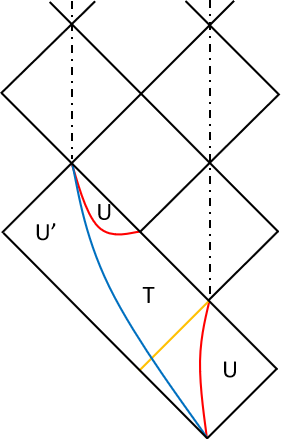}
\caption{\small There are two branches corresponding to Fig. \ref{av1} on the same plot. The final states of these two branches are both regular RN black holes. We set $Q=3.72, \gamma_1=4.0, \gamma_2=10.3, a_0=3.44$.}\label{RNra}
\end{figure}
\begin{figure}[h]
\centering
\includegraphics[width=7cm]{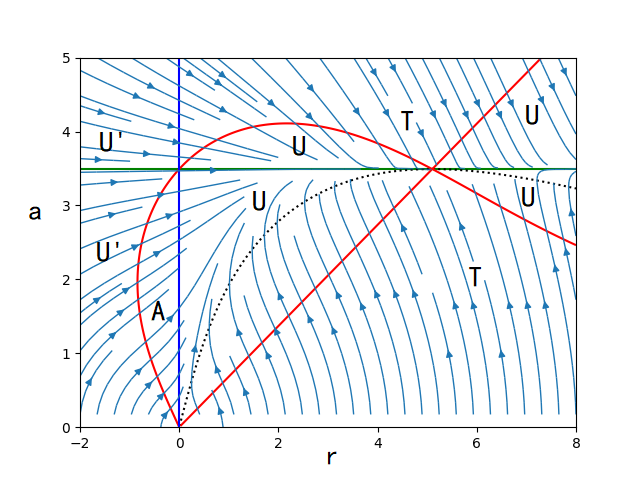}
\includegraphics[width=3cm]{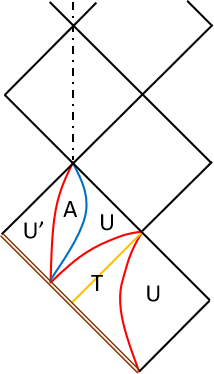}
\includegraphics[width=3cm]{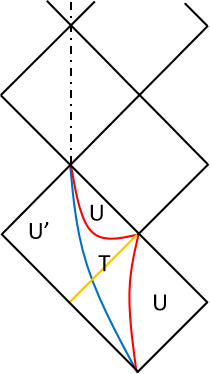}
\caption{\small In some cases, the final states can also be extreme black holes, where the inner horizon covers the outer horizon. We set $Q=3.72,\gamma_1=4.0,\gamma_2=10.3,a_0=3.49$.}\label{RNextrem}
\end{figure}

The upper branch of the $r-a$ plot in Fig. \ref{RNra} shows how a wormhole evolves into a black hole. The time direction is from up to down since decreasing $a$ means increasing $v$.
In the far past, the value of $a$ was large. There are two trapping horizons. 
Refer to the preceding definitions in Table I: 
both of them are FO type and inside the local lightcone. 
They are the wormhole horizons under the definition of Hayward.
According to \eqref{case1}, during the large $a$ era, we could ignore the log term.
The function $a(v)$ is almost a linear function of $v$ such that we could claim that the early state of the upper branch is similar to the evolving Ellis wormhole.  
As $a$ closes to the $a=a_0$, another part of the red line appears. 
It shows that a degenerate marginal surface, which is the top of the new part of the red line, satisfies $\mathcal{L}_l\theta_k=0$ and emerges between two trapping horizons as $v$ increases.
Then the degenerate marginal surface splits as two trapping horizons.
The left red one is again a FOTH while the right red one is a FITH.
The spacetime evolves to the quasi-black hole state with four trapping horizons. 
Finally, the quasi-black hole becomes a static RN-like black hole in the stationary point. 
The new red FOTH tends to $r=0$ and combines with the blue line to become a stationary wormhole throat, i.e., a minimum surface.
Meanwhile, the new red FITH tends to the Cauchy horizon corresponding to the inner horizon of the static RN-like black hole.
The old part of the red line, the rightmost FOTH, tends to the event horizon, which corresponds to the outer horizon.
The whole spacetime is not singular in this case.

The lower branch in the $r-a$ plot of Fig.\ref{RNra} shows how a null singularity collapses into a black hole by absorbing phantomlike negative energy (see Table II
). At the beginning of the evolution, namely $v=0$, there is a null singularity with divergent mass. The singularity changes to the quasi-black hole state quickly. 
The rightmost red contour is the FOTH and it tends to the event horizon.
It is temporal because it is inside the light cone.
The other three trapping horizons start from the center $r=0$.
The blue contour $r=0$ and the red one at its left are both POTH. 
They finally coalesce to become a wormhole throat, the minimum surface, located at $r=0$ as $v$ increases to positive infinity. 
The red contour just at the right of $r=0$ is FITH and tends to an inner horizon in the final state.

In the following discussion for other cases, we would use the same color with Fig.\ref{RNra} to label corresponding contours and streamlines in the $r-a$ plot and hypersurfaces in the Penrose diagram. 
When the ratio $Q/a$ equals the critical value $8(\gamma_1+\sqrt{\gamma_1^2+\gamma_2^2})/\gamma_2^2$, the extreme black hole would finally form. We plot this situation in Fig.\ref{RNextrem}.

\noindent{\bf Charged Ellis wormhole:}

In Fig.\ref{Ellis}, we recover the charged Ellis wormhole in Ref.\cite{Huang:2019arj} when the ratio $Q/a$ is smaller than $8(\gamma_1+\sqrt{\gamma_1^2+\gamma_2^2})/\gamma_2^2$. 
The regime connected with the rightmost side, bounded by the up and down red contours corresponds to $\theta_k>0$. The complement regime excluded from the red line itself corresponds to $\theta_k<0$.
Thus we could label regimes divided by the red and blue contours similar to the above discussion.
There are two paths to evolve to a charged Ellis wormhole. One path is from a null singularity to a wormhole. This case is shown in the lower branch of Fig.\ref{Ellis}.
We can see the red FITH and the red FOTH at the right side of $r=0$ touch each other at a sufficiently large $v$ moment.
Two POTH at the left become a minimum surface at the final state.
It has failed to form a black hole.
There is no event horizon since every streamline could go to the positive infinity of $r$.

\begin{figure}[h]
\centering
\includegraphics[width=6.1cm]{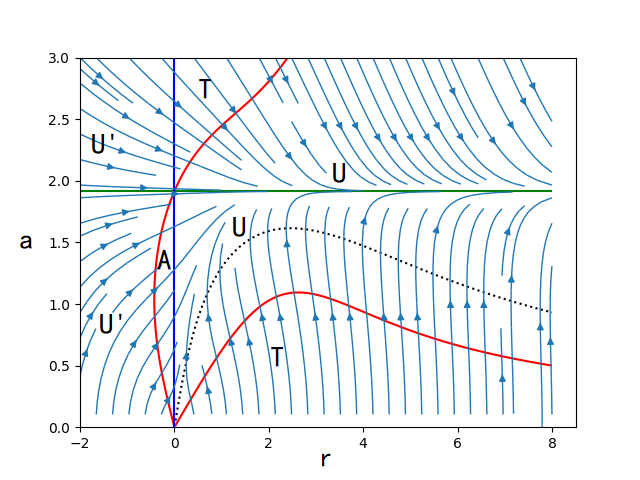}
\includegraphics[width=4.4cm]{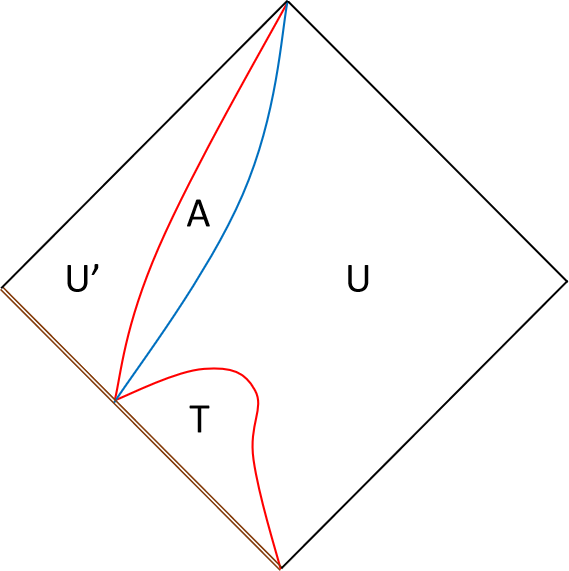}
\includegraphics[width=4.4cm]{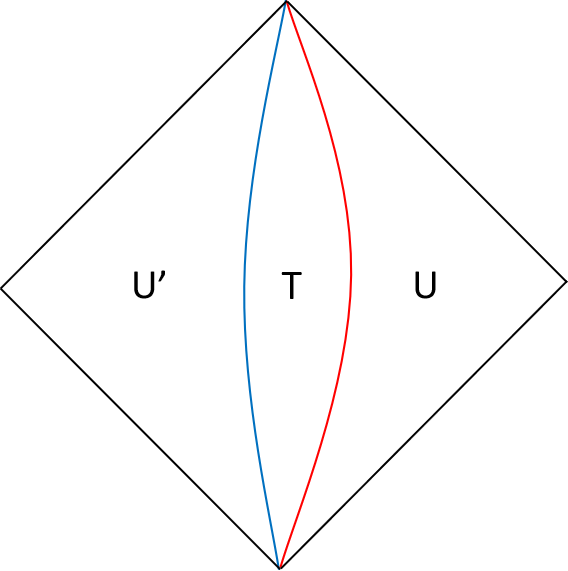}
\includegraphics[width=6.1cm]{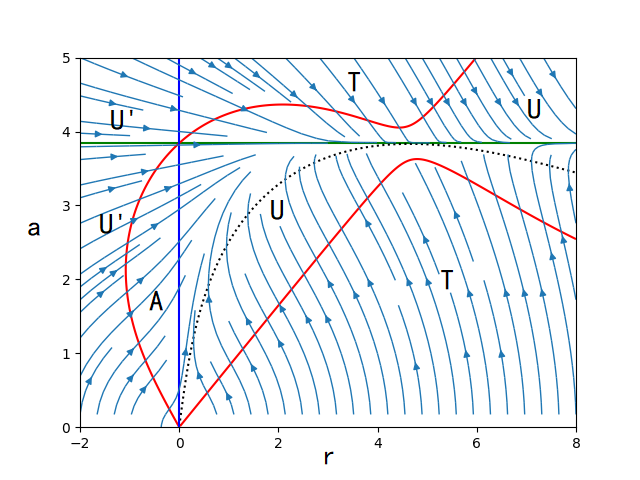}
\includegraphics[width=4.4cm]{plusnorootlower.png}
\includegraphics[width=4.4cm]{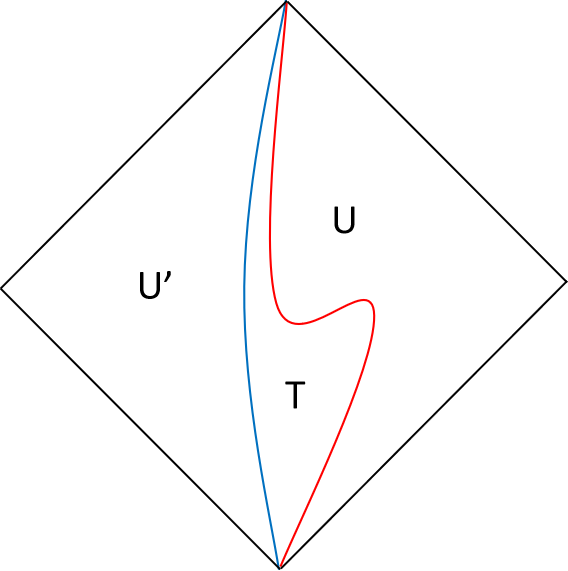}
\caption{\small These two plots show that the evolution leads to the charged Ellis wormhole in Ref.\cite{Huang:2019arj} as the final state. Despite the lower branch of two plots giving the similar evolution for the null singularity to the final wormhole, the upper branch of the lower plot describes a quasi-black hole forming during the evolution, but there is no quasi-black hole state in upper branch of the upper $r-a$ plot. We set $Q=3.72,\gamma_1=1,\gamma_2=2.3,a_0=1.92$ for the the upper plot and   $Q=3.72,\gamma_1=2.0,\gamma_2=10.3,a_0=3.84$ for the lower plot.}\label{Ellis}
\end{figure}

Another path is from the dynamic wormhole state to the static wormhole. 
The upper branch of the upper $r-a$ plot in Fig. \ref{Ellis} shows that the red FOTH experiences an area-decreasing era and finally coalesces with the blue FOTH to form a minimal surface. 
It is worth noting that the upper branch of the the lower $r-a$ plot is different.
It shows that a quasi-black hole fails to form an event horizon.
During the evolution, a degenerate marginal surface suddenly emerges and splits as a temporal FOTH at the left and a spatial FITH at the right.
In this era the moment $v$ is in the quasi-black hole state because the marginal surface near to the rightmost FOTH is FI type.
The FITH then contacts with the rightmost FOTH, becomes another degenerate marginal surface, and disappears.
The left red FOTH finally combines with the blue FOTH $r=0$ to form a minimal surface. 
All of the processes portrayed by Fig. \ref{Ellis} are about the formation of a traversable charged Ellis wormhole.

\subsubsection{$\gamma_1<0$}

Besides many similar cases of the RN-like black hole and the charged Ellis wormhole in $\gamma_1>0$, there is a new final state at $\gamma_1<0$. The new final state is nothing but the black bounce\cite{Simpson:2018tsi, Huang:2019arj}. 

\noindent{\bf Black bounce:}
The concept of black bounce is a bounce replacing the spatial singularity inside the black hole event horizon.  
It is a kind of regular black hole.
Researchers expect some quantum gravity effect would give the minimum size to prevent singularity forms and the black bounce may be a semi-classical geometric description for it.
Despite that the bounce happens at the highly dynamic regime, it is still a Killing vector field in those proposed semi-classical metrics.
But there is not a Killing vector in our dynamic solution for the corresponding cases.
\begin{figure}[h]
\centering
\includegraphics[width=7cm]{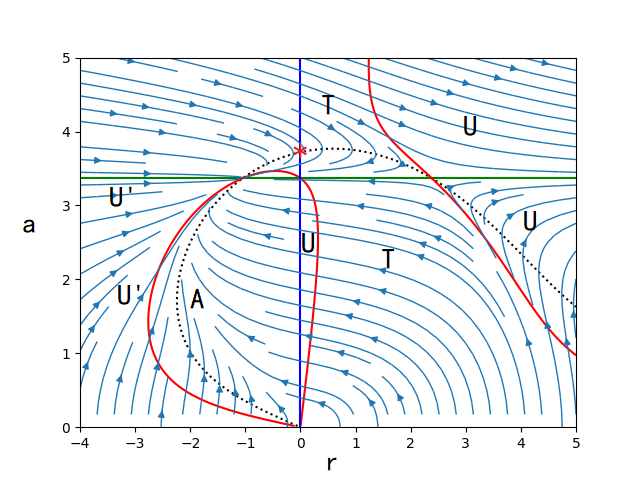}
\includegraphics[width=3.3cm]{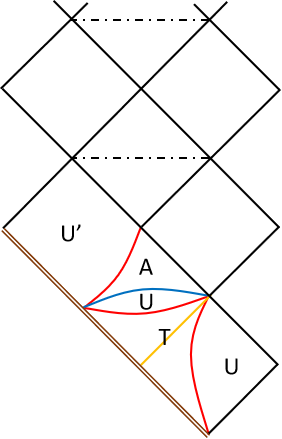}
\includegraphics[width=3.3cm]{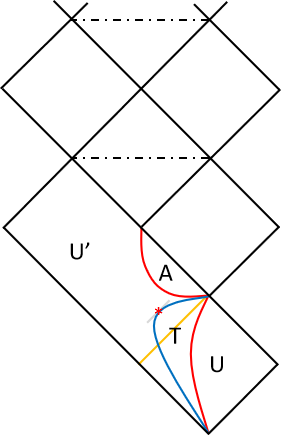}
\includegraphics[width=7cm]{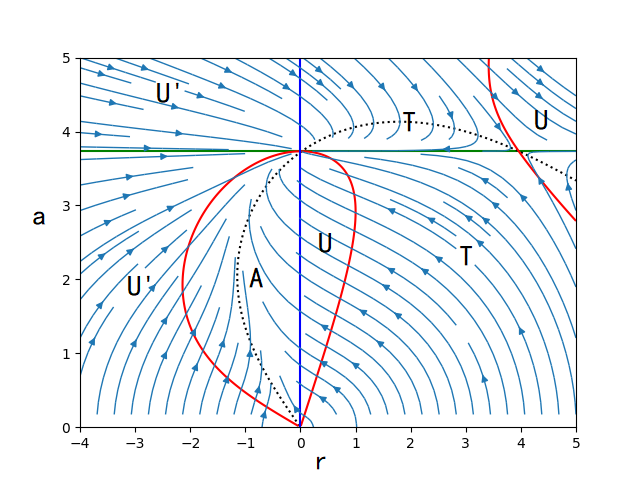}
\includegraphics[width=3.3cm]{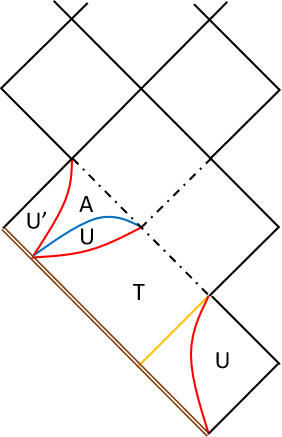}
\includegraphics[width=3.3cm]{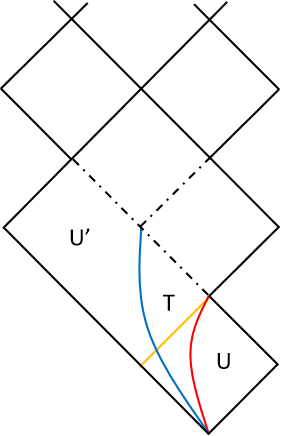}
\caption{\small The final states are black bounces. The upper $r-a$ plot shows the cases where the bounce happens between the black hole event horizon and the inner Cauchy horizon. In the lower plot, the bounce concides with the $r=r_-$. We set $Q=3.72,\gamma_1=-4.0,\gamma_2=1.3,a_0=3.36$ for the upper plot and  $Q=3.72,\gamma_1=-4.0,\gamma_2=4.3,a_0=3.73$ for the lower plot. }\label{blackbounce}
\end{figure}
Similar to the Fig.\ref{RNra}, both the right side of the rightmost red contour and the regime enclosed by the closed red contour correspond to $\theta_k>0$. Thus we could justify the divided regime for the plots in Fig.\ref{blackbounce}.

The lower branch of the upper $r-a$ plot in Fig.\ref{blackbounce} shows that a null singularity evolves to a black bounce. There are three trapping horizons near $r=0$.
The blue contour $r=0$ is a spatial PITH.
The red contour at the left of the $r=0$ is a temporal POTH, while the right one is a spatial FITH.
Along the streamlines through $r=0$, there is an untrapped regime bounded by the blue PITH and the red FITH. It connects the trapped regime through the red FITH and the antitrapped regime through the blue PITH.
Thus this configuration is the bounce defined by us in the last part of Sec. III .
These two spatial trapping horizons tend to the minimum surface. 
The leftmost red POTH tends to the $r_-$ which is the intersection of the $h=0$ and the $a=a_0$ at the left of the $r=0$.
The rightmost red contour is a temporal FOTH and tends to the $r_+$, the positive intersection of the $h=0$ and the $a=a_0$.
Again, with the help of the $h<0$ regime, we could justify that most of the streamlines leave the $r_+$.
Their fate is neither to escape to the positive infinity of $r$ nor to converge to the $r_-$.
The event horizon is formed by those streamlines which hit the $r_+$ exactly. It is easier to be seen in the Penrose diagram.
Since it is hard to say the bounce is static even with the Killing vector field, we avoid naming the static black bounce. 
Instead, we claim that the final geometry of the above evolution should be the black bounce with the Killing vector field.

The upper branch of the same plot with the previous shows that a dynamic wormhole state evolves to a black bounce. 
The blue temporal FOTH becomes a degenerate marginal surface at a particular moment.
We label a star mark at the intersection of the $r=0$ and the $h=0$ to represent the degenerate marginal surface.
After that, the blue contour $r=0$ becomes a FITH.
The rightmost red FOTH just tends to the $r_+$ again and marks that there is a set of streamlines to form the event horizon finally.
Interestingly, the top of the red contour at the left of the $r=0$ is another degenerate marginal surface.
It emerges after the $r=0$ changes its type and splits into two trapping horizons.
The left red temporal POTH tends to the $r_-$ while the right spatial PITH tends to coalesce with the blue FITH. 
The final state is the same as the lower branch.

The lower $r-a$ plot in Fig.\ref{blackbounce} is the critical situation of black bounce, satisfied by $r_-=0$.
It means the bounce hypersurface is also the inner horizon.
The red contour is indeed a temporal FOTH and tends to the $r_+$.
We could still classify the streamlines into three classes: escape to infinity, converge to the $r_-=0$, and form the event horizon.
Back to the black bounce solution with the Killing field, we could see the minimum surface of the bounce coincides with one Killing horizon. 
For the upper branch, the whole history is the dynamic wormhole because the blue $r=0$ above the green line $a=a_0$ is also a temporal FOTH.
For the lower branch, there are four trapping horizons initially.
Let us focus on three of them near the center.
The blue spatial PITH, the left red temporal POTH, and the right red spatial FITH; all of these three trapping horizons tend to the $r=r_-=0$.
The spacetime behind the $r=r_-=0$ should have Killing vector field such that the $r=r_-=0$ itself becomes a Killing horizon, as is shown in the Penrose diagram.

\noindent{\bf RN-like black hole and charged Ellis wormhole:}

In $\gamma_1<0$, we recover the RN-like black hole and the charged Ellis wormhole under the appropriate tuning of $Q$ and $a$. We show these situations in Fig. \ref{blackbouncetoRN2}, \ref{blackbouncetoRNext}, and \ref{blackbouncetoChEllisWH}.
It is shown that the lower branches are different with the cases in $\gamma_1>0$.
The initial blue $r=0$ is a spatial PITH rather than a temporal POTH.
It does affect the final state because the blue trapping horizon changes its type before $a=a_0$.
The star marks in the plots label the degenerate marginal surface which represents the type changing.

\begin{figure}[h]
\centering
\includegraphics[width=7cm]{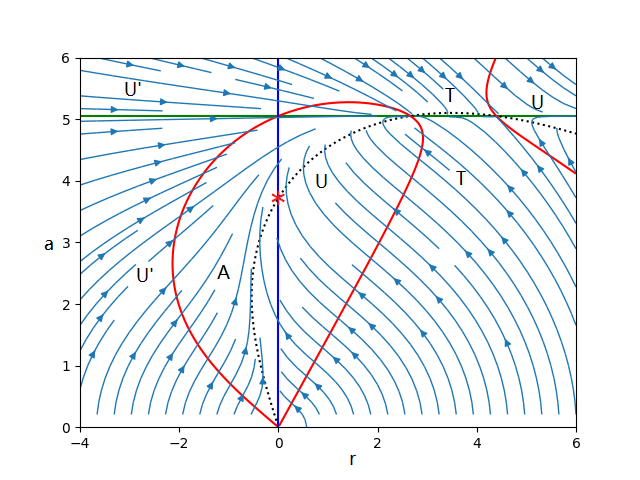}
\includegraphics[width=3.3cm]{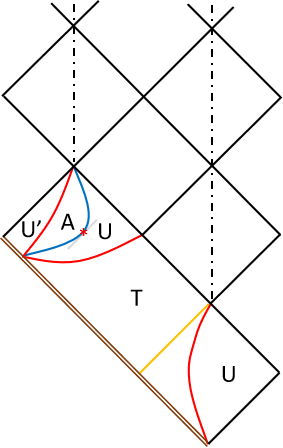}
\includegraphics[width=3.3cm]{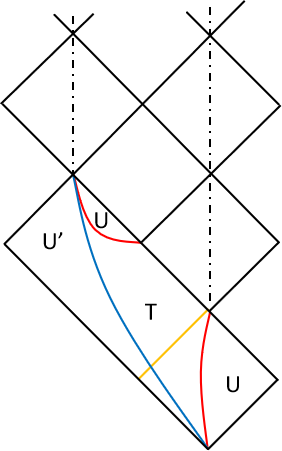}
\caption{\small The left one shows the final states are RN-like black holes under $\gamma_1<0$. We set $Q=3.72,\gamma_1=-4.0,\gamma_2=8.3,a_0=4.73$.}\label{blackbouncetoRN2}
\end{figure}

\begin{figure}[h]
\centering
\includegraphics[width=7cm]{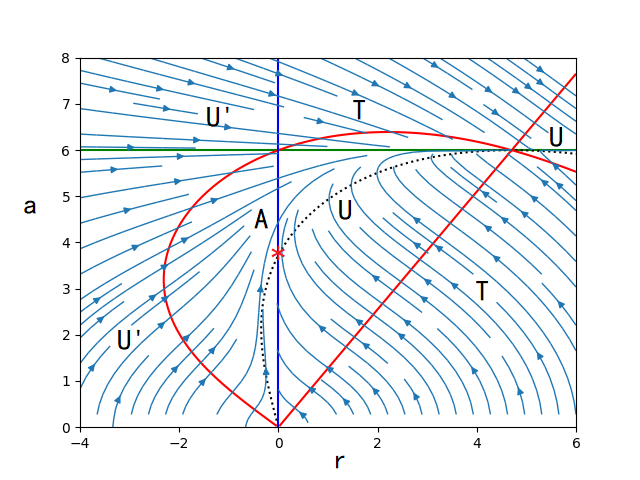}
\includegraphics[width=3.3cm]{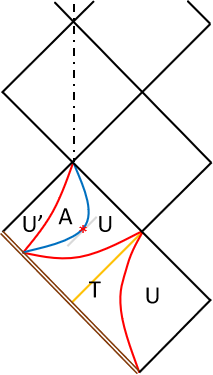}
\includegraphics[width=3.3cm]{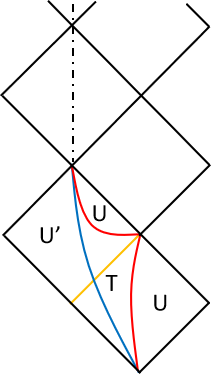}
\caption{\small The final states are extreme RN-like black holes under $\gamma_1<0$. We set $Q=3.72,\gamma_1=-4.0,\gamma_2=16.3,a_0=5.99$.}\label{blackbouncetoRNext}
\end{figure}

\begin{figure}[h]
\centering
\includegraphics[width=6.1cm]{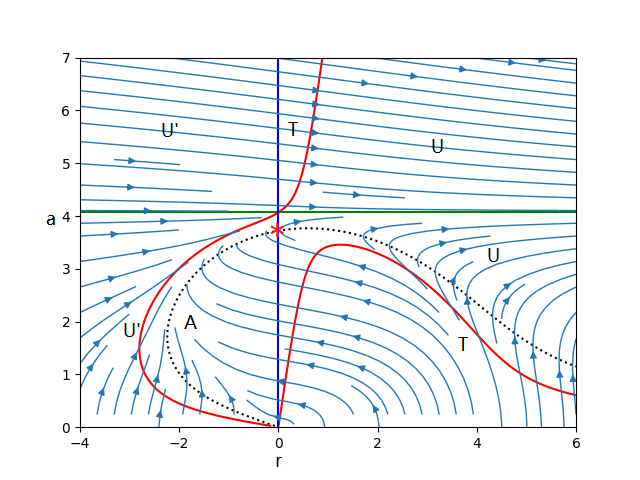}
\includegraphics[width=4.4cm]{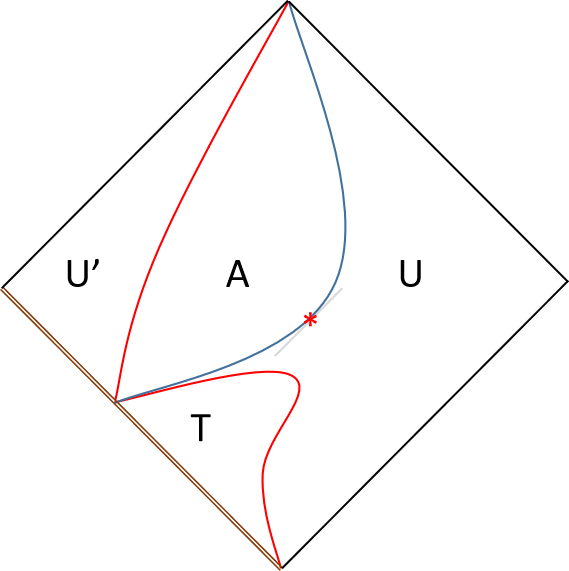}
\includegraphics[width=4.4cm]{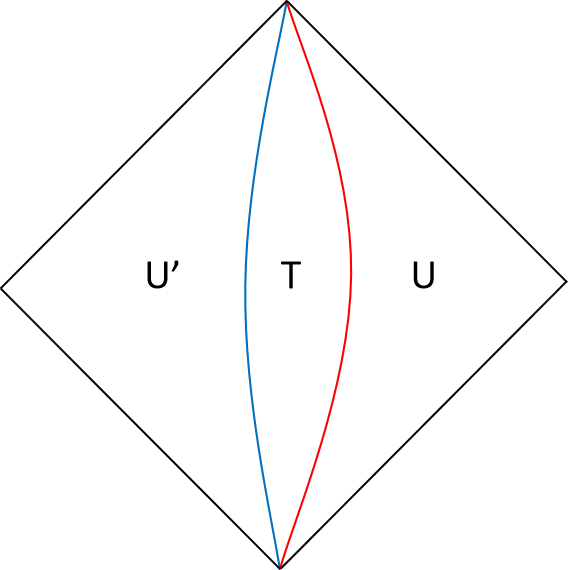}
\caption{\small The final states are charged Ellis wormholes. We set $Q=3.72,\gamma_1=-4.0,\gamma_2=1.3,a_0=4.07$.}\label{blackbouncetoChEllisWH}
\end{figure}

\subsection{Eternal evolution: Case II and case III}

Unlike the above section, case II and case III, i.e.,\eqref{evosluS} and \eqref{evosluF}, relate to the eternal evolution. We solve \eqref{evosluS} and \eqref{evosluF} directly and obtain
\be\label{evosluSs} 
  \dot{a}=\ft{Q^2\gamma_2}{16}\big( \ft{1}{a^2} + \ft{1}{a_0^2} \big),\qquad\Rightarrow\qquad
  \ft{a}{a_0}- \arctan{\ft{a}{a_0}} =\ft{Q^2\gamma_2}{16 a_0^3}(v-v_0),\,
\ee
\be\label{evosluFs}
 \dot{a}=\ft{Q^2\gamma_2}{16 a^2},\qquad \Rightarrow\qquad \ft{a^3}{3}=\ft{Q^2\gamma_2}{16}(v-v_0),
\ee
where $a(0)=0$ means both cases begin with a null singularity. After that, the singularity will evolve to a wormhole as $v\to\infty$. The scalar charge $a(v)$ is increasing monotonically as what we draw in Fig.\ref{av22}.
\begin{figure}[h]
\centering
\includegraphics[width=6cm]{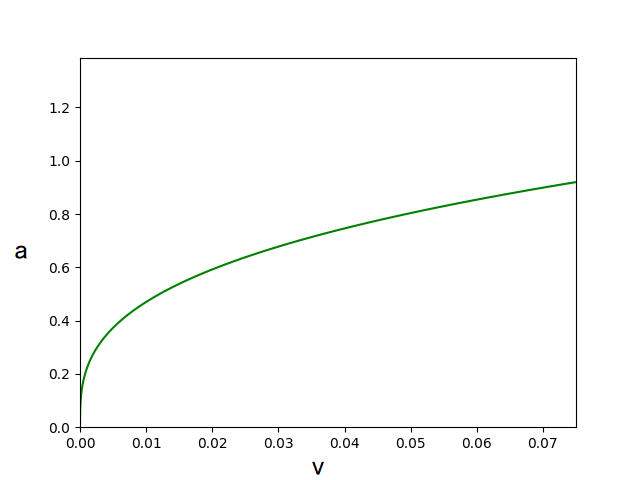}
\includegraphics[width=6cm]{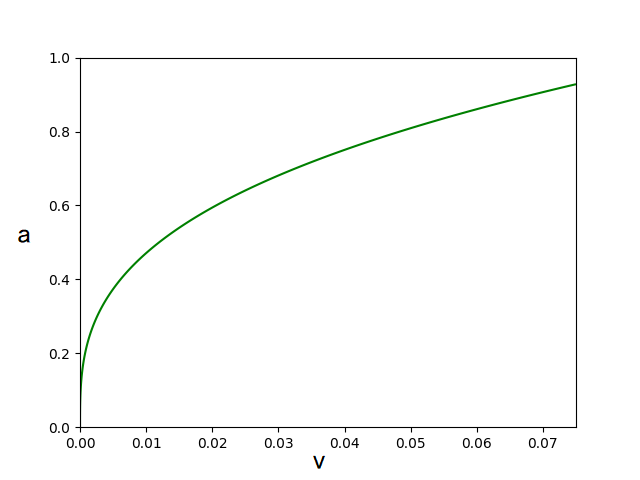}
\caption{\small In case II (left plot) and case III (right plot), the function $a(v)$ is always increased with time flow $v$. We set $Q=3.72, \gamma_1=-4.0, \gamma_2=4$, and $ a_0=4.39$. }\label{av22}
\end{figure}

Varieties of evolutionary processes happen in these cases. Although every case portrays the evolution from the initial null singularity to the final dynamic wormhole, the process could include multihorizons. To analyze these processes, we classify them into $\gamma_1>0$ and $\gamma_1<0$ again. Noting that case III shares similar properties with case II, we only represent case II.

\subsubsection{$\gamma_1>0$}

Figure.\ref{CaseIIgamma1plus} shows that two situations in $\gamma_1>0$ both start from a null singularity and then quickly change to a quasi-black hole. 
The red contour at the large positive $r$ is a temporal FOTH.
Near to the center, the blue contour is a temporal POTH and the left red one is again a temporal POTH while the right one is a spatial FITH.
Both of the plots show the same fate for the red FITH at $r>0$ and the rightmost FOTH.
They touch each other, become a degenerate marginal surface, and vanish.
Things are different at $r<0$.
In the upper $r-a$ plot of Fig. \ref{CaseIIgamma1plus}, the red POTH simply evolves to the left.
But in the lower plot, another red degenerate marginal surface emerges and splits into a spatial PITH at the right and a temporal POTH at the left.
The spatial PITH combines with the rightmost old POTH, forms a degenerate marginal surface, and disappears\footnote{More precisely, the moment intersected with four trapping horizons appears twice. The second one should be the quasi-white hole state in the point of view of the observers located at $r<0$.}.
Only the left POTH survives. After the disappearance of the degenerate marginal surface, the spacetime becomes a dynamic wormhole.

\begin{figure}[h]
\centering
\includegraphics[width=5cm]{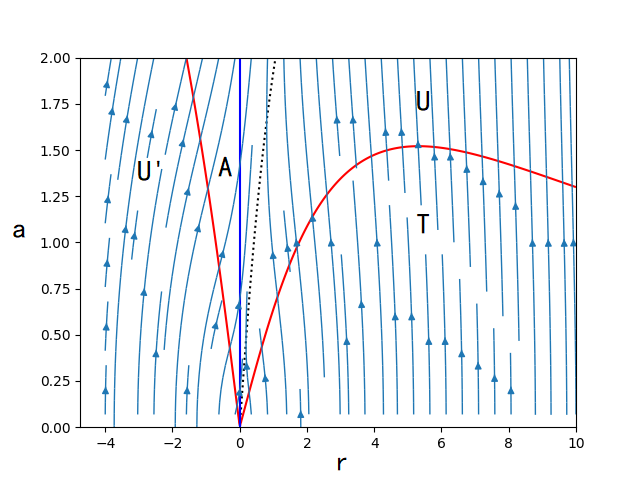}\qquad\qquad\qquad
\includegraphics[width=4cm]{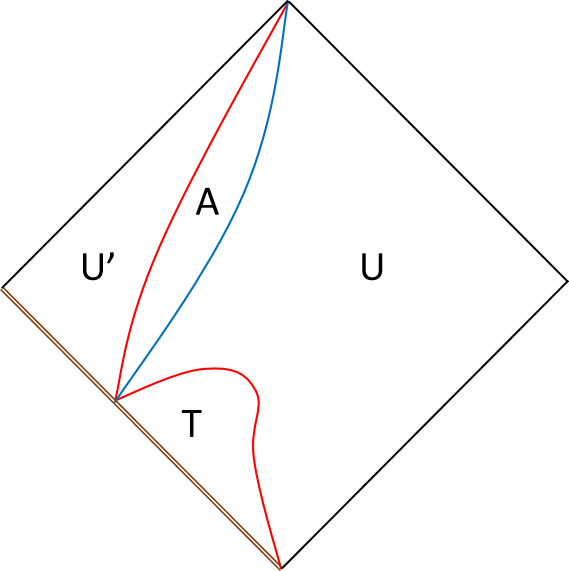}
\\

\includegraphics[width=5cm]{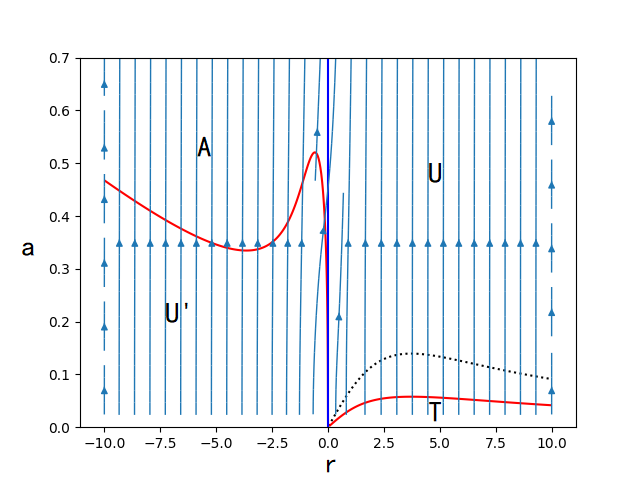}\qquad\qquad\qquad
\includegraphics[width=4cm]{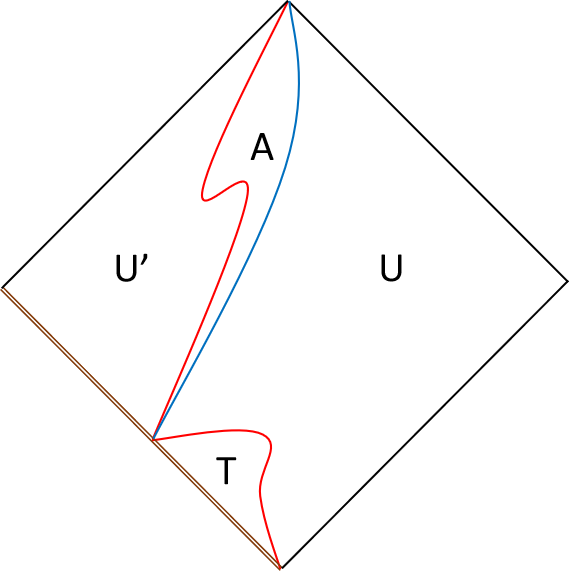}
\caption{\small Both of the plots shows that the singularity evolves to the dynamic wormhole. Especially, the lower one describes that the state intersected with four trapping horizons appears twice during the evolution. We set $Q=3.72,\gamma_1=4.0,\gamma_2=10.3,a_0=3.49$ for the upper plot and   $Q=3.72,\gamma_1=4.0,\gamma_2=0.3,a_0=0.14$ for the lower plot. }\label{CaseIIgamma1plus}
\end{figure}

\subsubsection{$\gamma_1<0$}
When $\gamma_1<0$, as is shown in Fig.\ref{CaseIIgamma1minus}, the initial $r=0$ is a spatial PITH.
Then it changes to a temporal POTH through a blue degenerate marginal surface labeled by the star mark.
The evolving picture for $r>0$ is the same with the $\gamma_1>0$ situation.
The initially born red spatial FITH combines with the rightmost red FOTH and vanishes together.
Two $r-a$ plots in Fig.\ref{CaseIIgamma1minus} show that the evolution of the left temporal red POTH is simple and just has a slight difference.
Compared with the case of $\gamma_1>0$, there is no emergence of any degenerate marginal surface.
We only draw one Penrose diagram for these two cases.
The whole picture is still the beginning as the null singularity evolves to a quasi-black hole and finally becomes a dynamic wormhole state.

\begin{figure}[h]
\centering
\includegraphics[width=5cm]{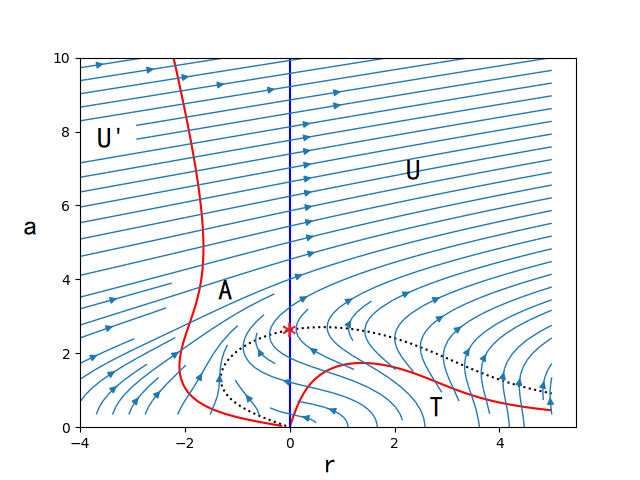}
\includegraphics[width=5cm]{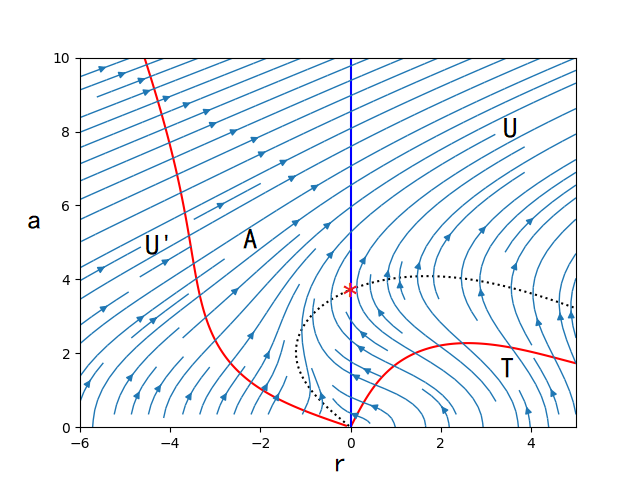}
\includegraphics[width=4cm]{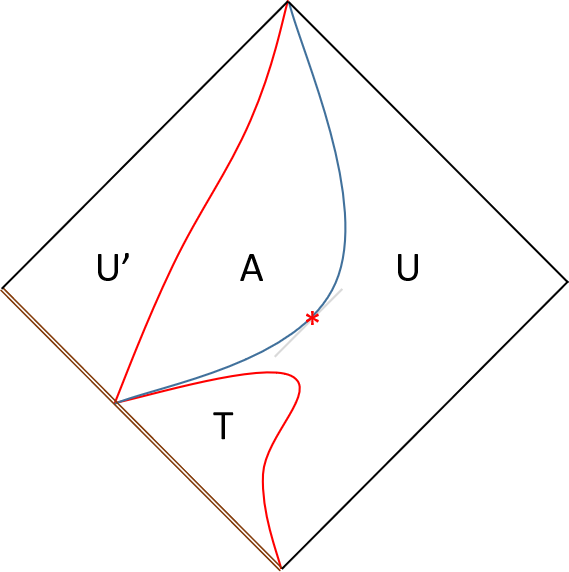}
\caption{\small Both of the plots shows that the singularity evolves to the dynamic wormhole. The quasi-black hole state appears just once during the evolution. We set $Q=3.72,\gamma_1=-2.0,\gamma_2=1, a_0=3.01$ for the left figure and $Q=3.72,\gamma_1=-4.0,\gamma_2=4.0,a_0=4.38$ for the right figure.}\label{CaseIIgamma1minus}
\end{figure}

\section{Conclusion and discussion}

We investigated the evolution of black holes and wormholes and their transition in this paper. Generalizing the static metric we constructed in Ref.\cite{Huang:2019arj}, we obtained a family of dynamic solutions in the Einstein-Maxwell-scalar theory. The solutions may depict various processes, especially including black hole/wormhole transitions, up to the parameters. 

To study dynamic black holes and wormholes, we used the concept of the ``trapping horizon" to follow their evolutions and transitions with each other. The trapping horizon is the hypersurface foliated by marginal surfaces. 
These definitions do not rely on the ``3+1" decomposition and are easily back to the static situations. We further introduced two concepts: ``quasi-black hole state" and ``dynamic wormhole state." They play important roles in describing middle states during the evolution.

In the dynamic solution we obtained above it is easy enough to track how trapping horizons evolve analytically. We listed all the situations of evolution. We found that a final static black hole or wormhole perhaps comes from a null singularity or a dynamic wormhole state. Furthermore, the black hole/wormhole transition may happen during the evolution.
We sketched the Penrose diagram to show the causal structures of the spacetime for every case.
For the situation with stationary point, though we have argued the $k^{\mu}$ geodesics could not stop at the $r_-$ and sketched the Penrose diagram, it is still worth it to study these geodesics by solving the equation $k^{\nu}\nabla_{\nu}k^{\mu} = \ft{h'}{2}k^{\mu}$ and drawing the Penrose diagram numerically. We have left it as our future research.

Dating back to Hayward's investigation, a trapping horizon obeys the so-called unified first law\cite{Hayward:1997jp}. 
For a dynamic black hole, the unified first law could be treated as the dynamic and quasilocal version of the first law for black hole thermodynamics.
Hayward's work indeed suggests wormholes also serve similar thermodynamic properties, as has been studied in Refs.\cite{Jamil:2009iu, Saiedi:2012qk, Debnath:2014vba, Rehman:2020myc}.
  
Another interesting feature is that our results show that a degenerate marginal surface could signal a trapping horizon's pair creation or annihilation. The surface comes out suddenly and evolves into one outer and one inner type of trapping horizon, or the time reversed process happens.
Since a slice of $a$ equated to some constant, which corresponds to an advanced time moment $v$, picks up a null rather than spatial hypersurface, it is hard to talk about space geometry at each moment.
It would be more helpful to chose other time slices such that every slice is a spatial hypersurface for constructing the evolving picture. 
One type of marginal surface emerges in space and splits into two marginal surfaces later.
One of them converses into other types of marginal surface, like inner to outer and vice versa.
The conversion is marked by the moment that the marginal surface becomes degenerate.

The above discussion inspires the picture that a degenerate marginal surface suddenly appears behind the horizon during the black hole evolution.
The black hole emitting Hawking radiation also has a temporal FOTH due to the decreasing of its area. If a degenerate marginal surface could emerge behind the black hole FOTH and split as one FOTH and one FITH, the FITH could evolve to contact the outer FOTH, become a degenerate marginal surface again, and disappear.
The inner FOTH still decreases its area. This picture could achieve a faster black hole horizon area decreasing.
Also, another semi-classical picture has been proposed in Refs. \cite{Hayward:2005gi, Binetruy:2018jfz}.
A similar feature is that the degenerate marginal surface suddenly appears or disappears. What they suggest is the picture that several trapping horizons glued by the degenerate surfaces form a closed hypersurface during the entire evolution of the black hole.
Does semi-classical quantum field theory support our picture? Which one is more plausible for the evolution of a black hole? We will leave these as open questions for future research.

\section*{Acknowledgements}

We are grateful to Hong L\"u, Mayumi Aoki, and Yuxuan Peng for useful discussions. Sijie Gao is highly appreciated, who advised us to discuss the evolution by using Penrose diagrams. J. Y. is supported by the
China Scholarship Council and the Japanese Government
(Monbukagakusho-MEXT) scholarship. H.H.~is supported by the Initial Research Foundation of Jiangxi Normal University.


\begin{thebibliography}{99}
 
	\bibitem{Abbott:2016blz}
	B.P.~Abbott \textit{et al.} (LIGO Scientific and Virgo Collaboration),
	\textit{Observation of gravitational waves from a binary black hole merger},
	Phys. Rev. Lett. \textbf{116}, 061102 (2016)
	doi:10.1103/PhysRevLett.116.061102,
	arXiv:1602.03837 [gr-qc].
	

	\bibitem{TheLIGOScientific:2016src}
	B.P.~Abbott \textit{et al.} (LIGO Scientific and Virgo Collaboration),
	\textit{Tests of general relativity with GW150914},
	Phys. Rev. Lett. \textbf{116}, 221101 (2016);
	Erratum, Phys. Rev. Lett. \textbf{121}, 129902(E) (2018),
	doi:10.1103/PhysRevLett.116.221101
	arXiv:1602.03841 [gr-qc].

	\bibitem{Akiyama:2019cqa}
	K.~Akiyama \textit{et al.} (Event Horizon Telescope Collaboration),
	\textit{First M87 event horizon telescope results. I. the shadow of the supermassive black hole},
	Astrophys. J. \textbf{875}, L1 (2019)
	doi:10.3847/2041-8213/ab0ec7
	arXiv:1906.11238 [astro-ph.GA].

   \bibitem{Penrose:1964wq} 
   R.~Penrose,
   \textit{Gravitational collapse and space-time singularities},
   Phys. Rev. Lett. {\bf 14}, 57(1965),
   doi:10.1103/PhysRevLett.14.57.
   
   \bibitem{Rees:1984si} 
   M.J.~Rees,
   \textit{Black hole models for active galactic nuclei},
   Ann. Rev. Astron. Astrophys. {\bf 22}, 471(1984),
   doi:10.1146/annurev.aa.22.090184.002351.
   
   \bibitem{Almheiri:2012rt} 
   A.~Almheiri, D.~Marolf, J.~Polchinski, and J.~Sully,
   \textit{Black holes: Complementarity or firewalls?},
   J. High Energy Phys. {\bf 02} (2013) 062,
   doi:10.1007/JHEP02(2013)062,
   arXiv:1207.3123 [hep-th].
   
   \bibitem{flamm}
   L.~Flamm,
   \textit{Comments on Einstein’s theory of gravity},
   Phys. Z. {\bf 17}, 448(1916).

   \bibitem{Einstein:1935tc}
   A.Einstein and N.~Rosen,
   \textit{The particle problem in the General Theory of Relativity},
   Phys. Rev. {\bf 48}, 73(1935),
   doi:10.1103/PhysRev.48.73.
   

   \bibitem{Morris:1988cz}
   M.S.~Morris and K.S.~Thorne,
   \textit{Wormholes in space-time and their use for interstellar travel: A tool for teaching general relativity},
   Am.\ J.\ Phys.\  {\bf 56}, 395 (1988).
   doi:10.1119/ 1.15620.
   
   \bibitem{Morris:1988tu}
   M.S.~Morris, K.S.~Thorne and U.~Yurtsever
   \textit{Wormholes, Time machines, and the Weak Energy Condition},
   Phys. Rev. Lett.  {\bf 61}, 1446 (1988).
   doi:10.1103/PhysRevLett.61.1446.
   

   \bibitem{Ellis:1973yv}
   H.G.~Ellis,
   \textit{Ether flow through a drainhole - a particle model in general relativity},
   J. Math. Phys. \textbf{14}, 104 (1973),
   doi:10.1063/1.1666161.

   \bibitem{Carroll:2003st} 
   S.M.~Carroll, M.~Hoffman, and M.~Trodden,
   \textit{Can the dark energy equation-of-state parameter $w$ be less than $-1$?},
   Phys. Rev. D {\bf 68}, 023509(2003),
   doi:10.1103/PhysRevD.68.023509,
   arXiv:astro-ph/0301273.
   

   \bibitem{Nojiri:2003vn} 
   S.~Nojiri and S.D.~Odintsov,
   \textit{Quantum de Sitter cosmology and phantom matter},
   Phys. Lett. B {\bf 562}, 147(2003),
   doi:10.1016/S0370-2693(03)00594-X,
   arXiv:hep-th/0303117.
   
  
   \bibitem{Bronnikov:1973fh}
   K.A.~Bronnikov,
   \textit{Scalar-tensor theory and scalar charge},
   Acta Phys. Polon. B \textbf{4}, 251 (1973).
  
   \bibitem{Ellis:1979bh} 
   H.G.~Ellis,
   \textit{The Evolving, flowless drain hole: A nongravitating particle model in General Relativity theory},
   Gen. Rel. Grav. \textbf{10}, 105(1979),
   doi:10.1007/BF00756794.
   


   \bibitem{Nozawa:2020wet} 
   M.~Nozawa,
   \textit{Static spacetimes haunted by a phantom scalar field II: Dilatonic charged solutions},
   Phys. Rev. D {\bf 103}, 024004 (2021),
   doi:10.1103/PhysRevD.103.024004,
   arXiv:2010.07560 [gr-qc].
   
   \bibitem{Chew:2018vjp}
   X.Y.~Chew, B.~Kleihaus, and J.~Kunz,
   \textit{Spinning wormholes in scalar-tensor Theory},
   Phys. Rev. D \textbf{97}, 064026 (2018),
   doi:10.1103/PhysRevD.97.064026,
   arXiv:1802.00365 [gr-qc].
   
   
   \bibitem{Chew:2019lsa}
   X.Y.~Chew, V.~Dzhunushaliev, V.~Folomeev, B.~Kleihaus, and J.~Kunz,
   \textit{Rotating wormhole solutions with a complex phantom scalar field},
   Phys. Rev. D \textbf{100}, 044019 (2019),
   doi:10.1103/PhysRevD.100.044019,
   arXiv:1906.08742 [gr-qc].
   
 
   \bibitem{Lazov:2017tjs} 
   B.~Lazov, P.~Nedkova, and S.~Yazadjiev,
   \textit{Uniqueness theorem for static phantom wormholes in Einstein\textendash{}Maxwell-dilaton theory},
   Phys. Lett. B {\bf 778}, 408(2018),
   doi:10.1016/j.physletb.2018.01.059,
   arXiv:1711.00290 [gr-qc].
   

   \bibitem{Lazov:2019bni} 
   B.~Lazov, P.~Nedkova, and Y.~Stoytcho,
   \textit{Uniqueness of static phantom wormhole solutions to the Einstein-Maxwell-dilaton equations},
   AIP Conf. Proc. {\bf 2075}, no.1, 090024(2019),
   doi:10.1063/1.5091238.
   

   \bibitem{Mai:2017riq} 
   Z.F.~Mai, and H.~Lu,
   \textit{Black holes, dark wormholes and solitons in f(T) gravities},
   Phys. Rev. D {\bf 95}, 124024 (2017),
   doi:10.1103/PhysRevD.95.124024,
   arXiv:1704.05919 [hep-th].
   
  
   \bibitem{Canate:2019spb} 
   P.~Ca\~nate, J.~Sultana, and D.~Kazanas,
   \textit{Ellis wormhole without a phantom scalar field},
   Phys. Rev. D {\bf 100}, 064007 (2019),
   doi:10.1103/PhysRevD.100.064007,
   arXiv:1907.09463 [gr-qc].
   

   \bibitem{Maldacena:2013xja} 
   J.~Maldacena, and L.~Susskind,
   \textit{Cool horizons for entangled black holes},
   Fortsch. Phys. {\bf 61}, 781(2013),
   doi:10.1002/prop.201300020,
   [arXiv:1306.0533(hep-th)].
   

   \bibitem{Gao:2016bin} 
   P.~Gao, D.L.~Jafferis, and A.C.~Wall
   \textit{Traversable wormholes via a double trace deformation},
   J. High Energy Phys. {\bf 12} (2017) 151,
   doi:10.1007/JHEP12(2017)151,
   arXiv:1608.05687 [hep-th].
   
  \bibitem{Maldacena:2020sxe}
   J.~Maldacena and A.~Milekhin,
   \textit{Humanly traversable wormholes},
   Phys. Rev. D \textbf{103}, 066007 (2021),
   doi:10.1103/PhysRevD.103.066007,
   arXiv:2008.06618 [hep-th].
   
   \bibitem{Cubrovic:2020iad}
   M.~\v{C}ubrovic,
   \textit{Fermions, hairy black holes and hairy wormholes in anti-de Sitter spaces},
   in 10th Mathematical Physics Meeting: School and Conference on Modern Mathematical Physics (2020).
   

   \bibitem{Maldacena:2018gjk}
   J.~Maldacena, A.~Milekhin, and F.~Popov,
   \textit{Traversable wormholes in four dimensions},
   arXiv:1807.04726 [hep-th].
  

   \bibitem{Blazquez-Salcedo:2019uqq}
   J.L.~Bl\'azquez-Salcedo, and C.~Knoll,
   \textit{Constructing spherically symmetric Einstein\textendash{}Dirac systems with multiple spinors: Ansatz, wormholes and other analytical solutions},
   Eur. Phys. J. C \textbf{80}, 174(2020);
   doi:10.1140/epjc/s10052-020-7706-3,
   arXiv:1910.03565 [gr-qc].
   
   \bibitem{Blazquez-Salcedo:2020czn}
   J.L.~Bl\'azquez-Salcedo, C.~Knoll, and E.~Radu
   \textit{Traversable Wormholes in Einstein-Dirac-Maxwell Theory},
   Phys. Rev. Lett. \textbf{126}, 101102(2021),
   doi:10.1103/PhysRevLett.126.101102,
   arXiv:2010.07317 [gr-qc].
     
   \bibitem{Simpson:2018tsi} 
   A.~Simpson, and M.~Visser,
   \textit{Black-bounce to traversable wormhole},
   J. Cosmol. Astropart. Phys. \textbf{02} (2019) 042,
   doi:10.1088/1475-7516/2019/02/042,
   arXiv:1812.07114 [gr-qc].
   
  
   \bibitem{Bronnikov:2006fu} 
   K.A.~Bronnikov, V.N.~Melnikov, and H.~Dehnen,
   \textit{Regular black holes and black universe},
   Gen. Rel. Grav. \textbf{39}, 973(2007),
   doi:10.1007/s10714-007-0430-6,
   arXiv:gr-qc/0611022.
   

   \bibitem{Ashtekar:2020ifw} 
   A.~Ashtekar,
   \textit{Black hole evaporation: a perspective from loop quantum gravity},
   Universe {\bf 6}, 21(2020),
   doi:10.3390/universe6020021,
   arXiv:2001.08833 [gr-qc].
   

   \bibitem{Tsukamoto:2020bjm}
   N.~Tsukamoto,
   \textit{Gravitational lensing in the Simpson-Visser black-bounce spacetime in a strong deflection limit},
   Phys. Rev. D \textbf{103}, 024033 (2021),
   doi:10.1103/PhysRevD.103.024033,
   arXiv:2011.03932 [gr-qc].
   

   \bibitem{Cheng:2021hoc}
   X.T.~Cheng, and Y.~Xie,
   \textit{Probing a black-bounce, traversable wormhole with weak deflection gravitational lensing},
   Phys. Rev. D \textbf{103}, 064040 (2021),
   doi:10.1103/PhysRevD.103.064040.
   

   \bibitem{Islam:2021ful}
   S.U.~Islam, J.~Kumar, and S.G.~Ghosh,
   \textit{Strong gravitational lensing by rotating Simpson--Visser black holes},
   arXiv:2104.00696 [gr-qc].
   

   \bibitem{Churilova:2019cyt} 
   M.S.~Churilova and Z.~Stuchlik,
   \textit{Ringing of the regular black-hole/wormhole transition},
   Classical Quantum Gravity {\bf 37}, 075014 (2020),
   doi:10.1088/1361-6382/ab7717,
   arXiv:1911.11823 [gr-qc].
   

   \bibitem{Bronnikov:2019sbx} 
   K.A.~Bronnikov, and R.A.~Konoplya,
   \textit{Echoes in brane worlds: Ringing at a black hole--wormhole transition},
   Phys. Rev. D {\bf 101}, 064004 (2020),
   doi:10.1103/PhysRevD.101.064004,
   arXiv:1912.05315 [gr-qc].
   

   \bibitem{Junior:2020lse} 
   H.C.D.~Lima, C.L.~Benone, and L.C.B~Crispino,
   \textit{Scalar absorption: Black holes versus wormholes},
   Phys. Rev. D {\bf 101}, 124009 (2020),
   doi:10.1103/PhysRevD.101.124009,
   arXiv:2006.03967 [gr-qc].

 
   \bibitem{Blazquez-Salcedo:2018ipc}
   J.L.~Bl\'azquez-Salcedo, X.Y.~Chew, and J.~Kunz,
   \textit{Scalar and axial quasinormal modes of massive static phantom wormholes},
   Phys. Rev. D \textbf{98}, 044035(2018),
   doi:10.1103/PhysRevD.98.044035,
   arXiv:1806.03282 [gr-qc].


\bibitem{Liu:2020qia}
H.~Liu, P.~Liu, Y.~Liu, B.~Wang, and J.~P.~Wu,
\textit{Echoes from phantom wormholes},
Phys. Rev. D \textbf{103}, 024006 (2021)
doi:10.1103/PhysRevD.103.024006,
arXiv:2007.09078 [gr-qc].




\bibitem{Bronnikov:2021liv}
K.~A.~Bronnikov, R.~A.~Konoplya, and T.~D.~Pappas,
\textit{General parametrization of wormhole spacetimes and its application to shadows and quasinormal modes},
Phys. Rev. D {\bf 103}, 124062 (2021),
doi:10.1103/PhysRevD.103.124062,
arXiv:2102.10679 [gr-qc].


   \bibitem{Simpson:2019cer} 
   A.~Simpson, P.~Martin-Moruno, and M.~Visser,
   \textit{Vaidya spacetimes, black-bounces, and traversable wormholes},
   Classical Quantum Gravity \textbf{36}, 145007 (2019),
   doi:10.1088/1361-6382/ab28a5,
   arXiv:1902.04232 [gr-qc].   



    \bibitem{Hayward:1998pp}
    S.A.~Hayward
    \textit{Dynamic wormholes},
    Int. J. Mod. Phys. D\textbf{8} 373 (1999),
    doi:10.1142/S0218271899000286,
    arXiv:gr-qc/9805019v3.   
    
    
   \bibitem{Hayward:2009yw}
    S.A.~Hayward
    \textit{Wormhole dynamics in spherical symmetry},
    Phys. Rev. D \textbf{79} 124001 (2009),
    doi:10.1103/PhysRevD.79.124001,
    arXiv:0903.5438 [gr-qc].
        
    
    
    \bibitem{Shinkai:2002gv}
    H.-a.~Shinkai and S.A.~Hayward,
    \textit{Fate of the first traversible wormhole: Black hole collapse or inflationary expansion},
    Phys. Rev. D \textbf{66}, 044005 (2002),
    doi:10.1103/PhysRevD.66.044005,
    arXiv:gr-qc/0205041.
    
   
    \bibitem{Mcnutt:2021qch} 
    D.D.~Mcnutt, W.~Julius, M.~Gorban, B.~Mattingly, P.~Brown, and G.~Cleaver
    \textit{Geometric surfaces: An invariant characterization of spherically symmetric black hole horizons and wormhole throats},
    Phys. Rev. D \textbf{103}, 124024 (2021),
    doi:10.1103/PhysRevD.103.124024,
    arXiv:2104.08935 [gr-qc].


    \bibitem{Cai:2008mh} 
    R.G.~Cai, L.M.~Cao, Y.P.~Hu, and S.P.~Kim,
    \textit{Generalized vaidya spacetime in Lovelock gravity and thermodynamics on apparent horizon},
    Phys. Rev. D \textbf{78}, 124012 (2008),
    doi:10.1103/PhysRevD.78.124012,
    arXiv:0810.2610 [hep-th].
    

    \bibitem{Fan:2016yqv} 
    Z.Y.~Fan, B.~Chen, and H.~L\"u,
    \textit{Global structure of exact scalar hairy dynamical black holes},
    J. High Energy Phys. {\bf 05} (2016) 170,
    doi:10.1007/JHEP05(2016)170,
    arXiv:1601.07246 [hep-th].
    
  
    \bibitem{Huang:2019lsl} 
    H.~Huang, Z.Y.~Fan, and H.~L\"u,
    \textit{Static and dynamic charged black holes},
    Eur. Phys. J. C {\bf 79}, 975 (2019),
    doi:10.1140/epjc/s10052-019-7477-x,
    arXiv:1908.07970 [hep-th].

	\bibitem{Hayward:1993wb}
	S.A.~Hayward,
	\textit{General laws of black-hole dynamics},
	Phys. Rev. D \textbf{49}, 6467 (1994), 
	doi:10.1103/PhysRevD.49.6467,
	arXiv:gr-qc/9303006.
	

	\bibitem{Hayward:1994bu}
	S.A.~Hayward,
	\textit{Gravitational energy in spherical symmetry},
	Phys. Rev. D \textbf{53}, 1938 (1996),
	doi:10.1103/PhysRevD.53.1938,
	arXiv:gr-qc/9408002v1.
	

	\bibitem{Hayward:1997jp}
	S.A.~Hayward,
	\textit{Unified first law of black-hole dynamics and relativistic thermodynamics},
	Classical Quantum Gravity \textbf{15}, 3147 (1998),
	doi:10.1088/0264-9381/15/10/017,
	arXiv:gr-qc/9710089v2.
	
	

	\bibitem{Hayward:2005gi}
	S.A.~Hayward,
	\textit{Formation and Evaporation of Regular Black Holes},
	Phys. Rev. Lett. \textbf{96} 031103 (2006),
	doi:10.1103/PhysRevLett.96.031103,
	arXiv:gr-qc/0506126.


    \bibitem{Hochberg:1998ii} 
    D.~Hochberg and M.~Visser,
    \textit{The Null Energy Condition in Dynamic Wormholes},
    Phys. Rev. Lett. {\bf 81}, 746 (1998),
    doi:10.1103/PhysRevLett.81.746,
    arXiv:gr-qc/9802048.
    

    \bibitem{Maeda:2009tk} 
    H.~Maeda, T.~Harada, and B.J.~Carr,
    \textit{Cosmological wormholes},
    Phys. Rev. D {\bf 79}, 044034 (2009),
    doi:10.1103/PhysRevD.79.044034,
    arXiv:0901.1153 [gr-qc].


    \bibitem{Tomikawa:2015swa} 
    T.~Tomikawa, K.~Izumi, and T.~Shiromizu,
    \textit{New definition of a wormhole throat},
    Phys. Rev. D {\bf 91}, 104008 (2015),
    doi:10.1103/PhysRevD.91.104008,
    arXiv:1503.01926 [gr-qc].
    

    \bibitem{Bittencourt:2017yxq} 
    E.~Bittencourt, R.~Klippert, and G.B.~Santos,
    \textit{Dynamical wormhole definitions confronted},
    Classical Quantum Gravity {\bf 35}, 155009 (2018),
    doi:10.1088/1361-6382/aace31,
    arXiv:1707.01078 [gr-qc].
    
  
  	\bibitem{Huang:2019arj} 
  	H.~Huang and J.~Yang,
  	\textit{Charged Ellis wormhole and black bounce},
  	Phys. Rev. D {\bf 100}, 124063 (2019),
  	doi:10.1103/PhysRevD.100.124063,
  	arXiv:1909.04603 [gr-qc].
    
    \bibitem{Huang:2020qmn}
    H.~Huang, H.~,L\"u and J.~Yang,
    \textit{Bronnikov-like wormholes in Einstein-Scalar gravity},
  	arXiv:2010.00197 [gr-qc].


    \bibitem{Ashtekar:2004cn} 
    A.~Ashtekar, and B.~Krishnan,
    \textit{Isolated and dynamical horizons and their applications},
    Living Rev. Rel. \textbf{7}, 10 (2004),
    doi:10.12942/lrr-2004-10,
    arXiv:gr-qc/0407042.
    	

  	\bibitem{Sherif:2019vvo} 
  	A.~Sherif, R.~Goswami, and S.~Maharaj,
  	\textit{Some results on cosmological and astrophysical horizons and trapped surfaces},
  	Classical Quantum Gravity {\bf 36}, 215001 (2019),
  	doi:10.1088/1361-6382/ab45bc,
  	arXiv:1905.02056 [gr-qc].
  	

  	\bibitem{Raviteja:2020fzt} 
  	K.~Raviteja and S.Gutti,
  	\textit{Aspects of marginally trapped and antitrapped surfaces in a $D$-dimensional evolving dust model},
  	Phys. Rev. D {\bf 102}, 024072 (2020),
  	doi:10.1103/PhysRevD.102.024072,
  	arXiv:2006.06468 [gr-qc].
  
    \bibitem{Helou:2015zma} 
  	A.~Helou,
  	\textit{Dynamics of the four kinds of trapping horizons and existence of Hawking radiation}, 
  	arXiv:1505.07371 [gr-qc]. 
  	
  	
  	\bibitem{Raychaudhuri:1953yv} 
  	A.~Raychaudhuri,
  	\textit{Relativistic cosmology. 1.},
  	Phys. Rev. {\bf 98}, 1123 (1955),
  	doi:10.1103/PhysRev.98.1123.
    	

    \bibitem{Cai:2006rs} 
    R.G.~Cai and L.M.~Cao,
    \textit{Unified first law and thermodynamics of apparent horizon in FRW universe},
    Phys. Rev. D {\bf 75}, 064008 (2007),
    doi:10.1103/PhysRevD.75.064008,
    arXiv:gr-qc/0611071.
    
 
  	\bibitem{Jamil:2009iu} 
  	M.~Jamil and M.~Akbar,
    \textit{Wormhole thermodynamics at apparent horizons},
  	arXiv:0911.2556 [hep-th].
  	
  	
  	\bibitem{Saiedi:2012qk} 
  	H.~Saiedi,
  	\textit{Thermodynamics of evolving Lorentzian wormholes at apparent horizon in f(R) theory of gravity},
  	Mod. Phys. Lett. A {\bf 27}, 1250220 (2012),
  	doi:10.1142/S0217732312502203,
  	arXiv:1409.2179 [gr-qc].
  	
  
  	\bibitem{Debnath:2014vba} 
  	U.~Debnath, M.~Jamil, R.~Myrzakulov, and M.Akbar,
  	\textit{Thermodynamics of evolving Lorentzian wormhole at apparent and event horizons},
  	Int. J. Theor. Phys. {\bf 53}, 4083 (2014),
  	doi:10.1007/s10773-014-2159-9,
  	arXiv:1202.1706 [physics.gen-ph].  	
  	
  	
  	\bibitem{Rehman:2020myc} 
  	M.~Rehman and K.~Saifullah,
  	\textit{Thermodynamics of dynamical wormholes},
  	J. Cosmol. Astropart. Phys. {\bf 06} (2021) 020,
  	doi:10.1088/1475-7516/2021/06/020,
  	arXiv:2001.08457 [gr-qc].    
    
    
     
  	\bibitem{Binetruy:2018jfz} 
  	P.~Bin\'etruy, A.~Helou, and F.~Lamy,
  	\textit{Closed trapping horizons without singularity},
  	Phys. Rev. D {\bf 98}, 064058 (2018),
  	doi:10.1103/PhysRevD.98.064058,
  	arXiv:1804.03912 [gr-qc].
      

\end{thebibliography}
\end{document}